\documentclass[twocolumn,amsmath,aps,superscriptaddress]{revtex4}
\usepackage{graphicx,amssymb,upgreek}
\usepackage{hyperref}
\graphicspath{{Figures/}}
\begin{document}
\newcommand\jcap{JCAP}
\newcommand{\be}{\begin{equation}}
\newcommand{\ee}{\end{equation}}
\newcommand{\bq}{\begin{eqnarray}}
\newcommand{\eq}{\end{eqnarray}}
\newcommand{\bsq}{\begin{subequations}}
\newcommand{\esq}{\end{subequations}}
\newcommand{\bc}{\begin{center}}
\newcommand{\ec}{\end{center}}
\newcommand {\R}{{\mathcal R}}
\newcommand{\al}{\alpha}
\newcommand\lsim{\mathrel{\rlap{\lower4pt\hbox{\hskip1pt$\sim$}}
    \raise1pt\hbox{$<$}}}
\newcommand\gsim{\mathrel{\rlap{\lower4pt\hbox{\hskip1pt$\sim$}}
    \raise1pt\hbox{$>$}}}

\title{Lagrangian description of cosmic fluids: Mapping dark energy into unified dark energy}

\author{V. M. C. Ferreira}
\email[Electronic address: ]{vasco.ferreira@astro.up.pt}
\affiliation{Departamento de F\'{\i}sica e Astronomia, Faculdade de Ci\^encias, Universidade do Porto, Rua do Campo Alegre s/n, 4169-007 Porto, Portugal}
\affiliation{Instituto de Astrof\'{\i}sica e Ci\^encias do Espa{\c c}o, Universidade do Porto, CAUP, Rua das Estrelas, 4150-762 Porto, Portugal}
\affiliation{Centro de Astrof\'{\i}sica da Universidade do Porto, Rua das Estrelas, 4150-762 Porto, Portugal}

\author{P. P. Avelino}
\email[Electronic address: ]{pedro.avelino@astro.up.pt}
\affiliation{Departamento de F\'{\i}sica e Astronomia, Faculdade de Ci\^encias, Universidade do Porto, Rua do Campo Alegre s/n, 4169-007 Porto, Portugal}
\affiliation{Instituto de Astrof\'{\i}sica e Ci\^encias do Espa{\c c}o, Universidade do Porto, CAUP, Rua das Estrelas, 4150-762 Porto, Portugal}
\affiliation{Centro de Astrof\'{\i}sica da Universidade do Porto, Rua das Estrelas, 4150-762 Porto, Portugal}
\affiliation{School of Physics and Astronomy, University of Birmingham, Birmingham, B15 2TT, United Kingdom}

\author{R. P. L. Azevedo}
\email[Electronic address: ]{rplazevedo@fc.up.pt}
\affiliation{Departamento de F\'{\i}sica e Astronomia, Faculdade de Ci\^encias, Universidade do Porto, Rua do Campo Alegre s/n, 4169-007 Porto, Portugal}
\affiliation{Instituto de Astrof\'{\i}sica e Ci\^encias do Espa{\c c}o, Universidade do Porto, CAUP, Rua das Estrelas, 4150-762 Porto, Portugal}
\affiliation{Centro de Astrof\'{\i}sica da Universidade do Porto, Rua das Estrelas, 4150-762 Porto, Portugal}

\date{\today}
\begin{abstract}
We investigate the appropriateness of the use of different Lagrangians to describe various components of the cosmic energy budget, discussing the degeneracies between them in the absence of nonminimal couplings to gravity or other fields, and clarifying some misconceptions in the literature. We further demonstrate that these  degeneracies are generally broken for nonminimal coupled fluids, in which case the identification of the appropriate on-shell Lagrangian may become  essential in order to characterize the overall dynamics. We then show that models with the same on-shell Lagrangian may have different proper energy densities and use this result to map dark energy models into unified dark energy models in which dark matter and dark energy are described by the same perfect fluid. We determine the correspondence between their equation of state parameters and sound speeds, briefly discussing the linear sound speed problem of unified dark energy models as well as a possible way out associated with the nonlinear dynamics.
\end{abstract}

\maketitle 

\section{\label{sec:Introduction}Introduction}

The detection of a Higgs-like particle \cite{Aad:2012tfa, Chatrchyan:2012xdj} reinforces the idea that scalar fields play a fundamental role in physics. In cosmology scalar fields are central to the primordial inflation paradigm \cite{1983PhLB..129..177L, 1999PhR...314....1L, 1999PhLB..458..209A, 2006PhLA..350..315D, 2010ARNPS..60...27A} and  potential candidates to explain the current accelerated expansion of the universe \cite{1998AJ....116.1009R, 1999ApJ...517..565P, 2001PhRvD..63j3510A, 2006IJMPD..15.1753C, 2007PhRvD..76h3511K} or even cold dark matter (CDM) \cite{2011PhLB..702..107G, 2017PhRvD..95d3541H, Urena-Lopez:2019kud} (see also \cite{2012PhR...513....1C,Bamba:2012cp, 2015PhR...568....1J, 2016arXiv160702979A} for recent reviews). More generally, scalar fields have also been proposed in the literature to unify primordial inflation and dark energy (DE) \cite{2015IJMPD..2430014H} or to account for the entire dark sector [DE and dark matter (DM)] \cite{2001PhLB..511..265K, 2002PhLB..535...17B, 2002PhRvD..66h1301P, 2002PhRvD..66d3507B, 2004PhRvL..93a1301S, 2005MPLA...20.2075C, 2007MPLA...22.2893B, 2010JCAP...05..012L, 2018PhRvD..98d3515F, 2018PhRvD..98j3520L, Boshkayev:2019qcx} (see also  \cite{2006PhRvL..97p1301L,2009PhRvD..79j3517B,2011PhRvD..83f3502D} for a unified description of primordial inflation, DE and DM).

It is well known that a minimally coupled scalar field in general relativity admits a perfect fluid description \cite{Madsen:1988ph}. Perfect fluids often provide a sufficiently general framework to model the source of the gravitational field. In particular, at cosmological scales (with homogeneity and isotropy being assumed) it is common to model the energy content of the Universe as a collection of perfect isentropic and irrotational fluids or, equivalently (under certain conditions, which we will consider in the present paper), as a collection of purely kinetic scalar fields \cite{2010PhRvD..81j7301A,2010PhRvD..81j3511U, 2014CQGra..31e5006P}. 

A number of action functionals, corresponding to at least three different on-shell Lagrangians ($\mathcal{L}_{\rm on-shell}=-\rho$, $p$ or $T$, where $\rho$, $p$, and $T$ represent, respectively, the proper density, the proper pressure, and the trace of the energy-momentum tensor of the fluid), have been shown to define the dynamics of a perfect fluid \cite{PhysRevD.2.2762, doi:10.1063/1.1665861, 1977AnPhy.107....1S, 1978AnRFM..10..301T, Matarrese:1984zw, Brown:1992kc, Andersson:2006nr, 2012PhRvD..86h7502M, Avelino2018, Avelino2018a}. Although some of these models may be used to describe the same physics in the context of general relativity, in general this degeneracy is broken in the presence of a nonminimal coupling (NMC) to gravity \cite{Bertolami:2007gv,Bertolami:2008ab,Sotiriou:2008it,Faraoni:2009rk,Harko:2010zi,Bertolami:2010cw,Ribeiro2014,Azizi:2014qsa,Bertolami2014} or to the other fields \cite{1982PhRvD..25.1527B,Sandvik:2001rv,Anchordoqui:2003ij,Copeland:2003cv,Lee:2004vm,Koivisto:2005nr,Avelino:2008dc,Ayata:2012,Pourtsidou:2013nha,Faraoni:2014vra,Boehmer:2015kta,Boehmer:2015sha,Barros:2019rdv,Kase:2019veo}. Therefore, in these theories the identification of the correct form of the on-shell Lagrangian can be essential in order to extract meaningful predictions \cite{Azevedo2018a,Azevedo2019a}.

Here, we will explore the degeneracies between the energy-momentum tensor of a perfect fluid and the corresponding on-shell Lagrangian. We shall use them to establish a correspondence between DE and unified dark energy (UDE) models, clarifying some misconceptions in the literature. The outline of this paper is as follows.  In Sec. \ref{sec:Perfect fluid lagrangian} we start by considering several different models for a perfect fluid, discussing the degeneracies between them, in the absence of a NMC to gravity or other fields, and the appropriateness of the use of the corresponding Lagrangians to describe different components of the cosmic energy budget. In Sec. \ref{sec:NMC} we present several examples featuring a NMC between DE or gravity with the matter or radiation fields, and showcase the importance of the use of the appropriate on-shell Lagrangian. In Sec. \ref{sec:Model} we define a mapping between DE models described by purely kinetic Lagrangians and UDE models. We also characterize the correspondence between their equation of state and sound speed parameters, briefly discussing the linear sound speed problem of UDE models and a possible way out associated with the nonlinear dynamics. Finally, we conclude in Sec. \ref{sec:conc}.

Throughout this paper we use units such that $8\pi G=c=k_B=1$, where $G$ is Newton's gravitational constant, $c$ is the value of the speed of light in vacuum and $k_B$ is the Boltzmann constant. We also adopt the metric signature $(-,+,+,+)$. The Einstein summation convention will be used whenever a Greek or a Latin index variable appears twice in a single term, once in an upper (superscript) and once in a lower (subscript) position. 

\section{\label{sec:Perfect fluid lagrangian} Perfect Fluid Lagrangian Descriptions}

Consider a fluid characterized by the following intensive variables, defined in the local comoving inertial frame: the proper particle number density $n$, energy density $\rho$, isotropic pressure $p$ and entropy per particle $s$ \cite{1996astro.ph..9119M}. Also, assume that there are no creation or annihilation processes, so that the particle number is conserved (or equivalently $n \propto V^{-1}$, where $V$ is the physical volume). In this case, the local form of the first law of thermodynamics may be written as
\begin{equation}\label{eq: 1st law}
d\left(\frac{\rho}{n}\right)=-pd\left(\frac{1}{n}\right)+Td s\,.
\end{equation}
In the case of an isentropic flow, the entropy per particle is conserved and, consequently, Eq. \eqref{eq: 1st law} simplifies to
\begin{equation}\label{eq: 1st law isentropy}
d\left(\frac{\rho}{n}\right)=-pd\left(\frac{1}{n}\right)\,.
\end{equation}
Defining an equation of state $\rho=\rho\left(n\right)$ and solving Eq. (\ref{eq: 1st law isentropy}) with respect to $p$ leads to
\begin{equation}\label{eq: 1st law rho n}
p(n) = \mu n - \rho\left(n\right) \,,
\end{equation}
where $\mu = d\rho/dn$ is the chemical potential. On the other hand, if $p=p\left(n\right)$ is given then Eq. \eqref{eq: 1st law isentropy} implies that
\begin{equation}\label{eq: 1s law p n}
\rho(n) = mn +n\int^{n}\frac{p\left(n'\right)}{n^{'2}}dn' \,,
\end{equation}
where $m$ is an integration constant.

\subsection{Model I \label{subsec: model I}}

The derivation of the equations of motion of a perfect fluid from an action functional has been studied by several authors \cite{1978AnRFM..10..301T, PhysRevD.2.2762, 1977AnPhy.107....1S, doi:10.1063/1.1665861, Matarrese:1984zw, Brown:1992kc}. Here we shall consider a model described by the action (see, e.g. \cite{Brown:1992kc})
\begin{equation}\label{eq:Action PF irrotational isentropic}
S=\int d^{4}x\sqrt{-g} \, \mathcal{L}(g_{\alpha\beta}, j^\alpha, \phi)\,,
\end{equation}
where
\begin{equation}\label{eq:Lagrangian PF irrotational isentropic}
\mathcal{L}=F\left(|{\bf j}|\right)+j^{\alpha}\nabla_{\alpha}\phi\,,
\end{equation}
$g=\text{det}\left(g_{\alpha\beta}\right)$, $g_{\alpha\beta}$ are the components of the metric tensor, $j^{\alpha}$ are the components of a timelike vector field ${\bf j}$, $\phi$ is a scalar field, $F$ is a function of $|\bf j|$, and 
\begin{equation}
|{\bf j}|=\sqrt{-j^\alpha j_\alpha}\,. \label{ndef}
\end{equation}

Varying the action with respect to $j^\alpha$ and $\phi$ one obtains the following equations of motion 
\begin{eqnarray}\label{eq: 4-velocity potential}
\frac{\delta S}{\delta j^{\alpha}}=&0&=-\frac{1}{|{\bf j}|}\frac{dF}{d|{\bf j}|} j_\alpha+\nabla_{\alpha}\phi\,,\\
\label{eq: particle number conservation}
\frac{\delta S}{\delta\phi}=&0&=\nabla_{\alpha}j^{\alpha}\,.
\end{eqnarray}

The energy-momentum tensor is given by
\begin{equation}\label{eq: EMT definition}
T^{\alpha\beta}=\frac{2}{\sqrt{-g}}\frac{\delta\left(\sqrt{-g}\mathcal{L}\right)}{\delta g_{\alpha\beta}}=2\frac{\delta\mathcal{L}}{\delta g_{\alpha\beta}}+\mathcal{L}g^{\alpha\beta}\,.
\end{equation}
Substituting the Lagrangian defined in Eq. \eqref{eq:Lagrangian PF irrotational isentropic} into  Eq. \eqref{eq: EMT definition} and using Eq. \eqref{eq: 4-velocity potential}, one obtains
\begin{equation}\label{eq: EMT with F}
T^{\alpha\beta}=- \frac{dF}{d|{\bf j}|}\frac{j^{\alpha}j^{\beta}}{|{\bf j}|}+\left(F - |{\bf j}|\frac{dF}{d|{\bf j}|}\right)g^{\alpha\beta}\,.
\end{equation}
Once the following identifications are made:
\begin{eqnarray}\label{number n}
n&=&|{\bf j}|\,,\\
\label{energy F(n)}
\rho\left(n\right)&=&-F\,,\\
p\left(n\right)&=&F-n \frac{dF}{dn}\,, \label{pressure}\\
u^\alpha &=&\frac{j^\alpha}{n}  \label{4velocity}\,,
\end{eqnarray}
the energy-momentum tensor may be written in a perfect fluid form 
\begin{equation}\label{eq: EMT perfect fluid}
T^{\alpha\beta}=\left(\rho+p\right)u^{\alpha}u^{\beta}+pg^{\alpha\beta}\,,
\end{equation}
where $\rho$ and $p$ are the proper density and pressure, and $u^\alpha$ are the components of the 4-velocity (satisfying $u^\alpha u_\alpha=-1$). With the identifications made above Eq. \eqref{eq: 4-velocity potential} now defines the 4-velocity of the fluid
\begin{equation}\label{eq: 4-velocity scalar field}
u^{\alpha}=-\frac{\nabla^{\alpha}\phi}{\mu}\,,
\end{equation}
associated with an irrotational flow (meaning that the spatial components of $u^\alpha$ are curl-free in the local comoving inertial frame) while Eq. \eqref{eq: particle number conservation} represents the particle number conservation equation. Note that the condition $u^\alpha u_\alpha=-1$ implies that
\begin{equation}
\mu^{2}=2X\,,
\end{equation}
where
\begin{equation}
X \equiv -\frac12\nabla^{\alpha}\phi\nabla_{\alpha}\phi> 0\,.
\end{equation}
On the other hand, Eq. \eqref{eq: 1st law rho n} may be obtained from Eqs. \eqref{energy F(n)} and \eqref{pressure}, thus implying that the Lagrangian given in Eq. \eqref{eq:Lagrangian PF irrotational isentropic} describes an isentropic flow satisfying 
\begin{equation}
\nabla_{\alpha}(s j^\alpha)=0 \label{isentropic}\,.
\end{equation}
Since the entropy per particle $s$ is not a dynamical variable of our model, Eq. \eqref{isentropic} is, in this case, equivalent to the particle number conservation equation given in Eq. \eqref{eq: particle number conservation}.

\subsection{Model II \label{sec: modelII}}

Using Eqs. \eqref{number n}, \eqref{energy F(n)}, \eqref{pressure}, (\ref{4velocity}) and \eqref{eq: 4-velocity scalar field}, it is possible to show that the on-shell Lagrangian, defined off-shell in Eq. \eqref{eq:Lagrangian PF irrotational isentropic}, is equal to
\begin{equation}
\mathcal{L}_{\rm on-shell}=-\rho+n\frac{d \rho}{dn}=p\label{lagrangian1}\,.
\end{equation}
If $\mu\left(n\right)$ is a strictly monotonic function of $n$ (such that there is a one-to-one relation between $\mu$ and $n$) Eq. \eqref{eq: 1st law rho n} may be written as
\begin{equation}\label{eq: 1st law legendre transformation}
p\left(\mu\right)=\mu n -\rho\,,
\end{equation}
where $p\left(\mu\right)$ is the Legendre transform of $\rho\left(n\right)$. The conjugate variables are related through
\begin{equation}\label{eq: conjugate variables}
n= \frac{dp}{d\mu}\,,\quad\mu=\frac{d\rho}{dn}\,.
\end{equation}
Taking into account that $\mu=\pm\sqrt{2X}$ and assuming $\mu>0$ one finally obtains
\begin{equation}
n(X)= \frac{d X}{d\mu}p_{,X}=\sqrt{2X}p_{,X}\,.
\end{equation}
where a comma denotes a partial derivative (e.g., $p_{,X}\equiv dp/dX$). In combination with Eq. \eqref{lagrangian1} this implies that the pure $k$-essence Lagrangian $\mathcal{L}\left(X\right)=p\left(X\right)$ may be used to describe an irrotational perfect fluid with conserved particle number and constant entropy per particle \cite{2010PhRvD..81j7301A,2010PhRvD..81j3511U, 2014CQGra..31e5006P}. 

The equation of motion of the scalar field
\begin{equation}\label{eq: k-essence eq. of motion}
\nabla_{\alpha}\left(\mathcal{L}_{,X}\nabla^{\alpha}\phi\right) = 0
\end{equation}
provides the equivalent in the scalar field theory of the particle number conservation, given by Eq. \eqref{eq: particle number conservation}. Interestingly, the identifications $\mathcal{L}=p$, $u_\alpha=-\nabla_\alpha \phi/{\sqrt{2X}}$, in combination with $\rho = 2X\mathcal{L}_{,X}-\mathcal{L}$ are also required in order that the energy-momentum tensor
\begin{equation}\label{eq: EMT k-essence X}
T^{\alpha\beta}=\mathcal{L}_{,X}\nabla^{\alpha}\phi\nabla^{\beta}\phi+\mathcal{L}g^{\alpha\beta}\,,
\end{equation}
associated with an arbitrary scalar field Lagrangian $\mathcal{L}(\phi,X)$ may be written in a perfect fluid form.

\subsection{Model III \label{sec: modelIII}}

The transformation 
\begin{equation}
\mathcal{L} \to \mathcal{L} - \nabla_\alpha (\phi j^\alpha) \label{lagtransf}
\end{equation}
leaves the action in Eq. \eqref{eq:Action PF irrotational isentropic} unchanged up to surface terms. This implies that the equations of motion given in Eqs. \eqref{eq: 4-velocity potential} and \eqref{eq: particle number conservation} are insensitive to this transformation. The resulting off-shell Lagrangian is given by
\begin{eqnarray}
\mathcal{L}&=&F(n)+j^\alpha \nabla_\alpha \phi - \nabla_\alpha (\phi j^\alpha)\nonumber \\
&=&F(n)- \phi \nabla_\alpha j^\alpha\,. \label{transfL}
\end{eqnarray}
Varying the matter action with respect to the metric components one obtains
\begin{eqnarray}
\label{eq: EMT0}
\delta S&=&\int d^{4}x \frac{\delta\left(\sqrt{-g}\mathcal{L}\right)}{\delta g_{\alpha\beta}} \delta g_{\alpha \beta}\nonumber\\
&=&\frac12 \int d^{4}x {\sqrt{-g}} \, T^{\alpha \beta} \delta g_{\alpha \beta}\,,
\end{eqnarray}
where
\begin{eqnarray}\label{eq: EMT1}
\delta\left(\sqrt{-g}\mathcal{L}\right)&=& \sqrt{-g} \delta \mathcal{L} + \mathcal{L} \delta \sqrt{-g} \nonumber \\
&=& \sqrt{-g} \delta \mathcal{L} +  \frac{\mathcal{L}}{2}  \sqrt{-g} g^{\alpha\beta} \delta g_{\alpha\beta}\,,
\end{eqnarray}
with
\begin{equation}\label{eq: EMT2}
\delta \mathcal{L} =- \frac{1}{2}\frac{dF}{d|{\bf j}|}\frac{j^{\alpha}j^{\beta}}{|{\bf j}|} \delta g_{\alpha\beta} - \phi \delta (\nabla_\nu j^\nu) \,,
\end{equation}
and
\begin{eqnarray}\label{eq: EMT3}
\phi \delta \left(\nabla_\nu j^\nu\right)&=&\phi \delta \left(\frac{\partial_{\nu}\left(\sqrt{-g} j^{\nu}\right)}{\sqrt{-g}} \right) \nonumber \\
&=& -\frac12 g^{\alpha\beta} \delta g_{\alpha\beta} \nabla_\nu \left(\phi j^\nu\right)\nonumber\\
&+& \frac12\nabla_\nu\left( \phi j^\nu g^{\alpha\beta} \delta g_{\alpha\beta}\right)\,.
\end{eqnarray}
Discarding the last term in Eq. \eqref{eq: EMT3} --- this term gives rise to a vanishing surface term in Eq. \eqref{eq: EMT0} ($\delta g_{\alpha \beta}=0$ on the boundary) --- and using Eqs. \eqref{eq: 4-velocity potential} and \eqref{eq: particle number conservation} it is simple to show that the energy-momentum tensor associated with the transformed Lagrangian defined in \eqref{transfL} is still given by Eq. \eqref{eq: EMT with F}. However, in this case the on-shell Lagrangian is equal to
\begin{equation}
\mathcal{L}_{\rm on-shell}=F=-\rho\,.
\end{equation}
Using this result, in combination with Eq. \eqref{eq: 1s law p n}, it is possible to write the on-shell Lagrangian as
\begin{equation}\label{eq: Harko Lagrangian}
\mathcal{L}_{\rm on-shell}=-mn - n\int^{n}\frac{p\left(n'\right)}{n^{'2}}dn'
\end{equation}
(see also \cite{2012PhRvD..86h7502M} for an alternative derivation of this result).

\subsection{Model IV}

A more general Lagrangian often considered in the literature to describe a perfect fluid is given by \cite{Brown:1992kc} (see also \cite{Boehmer:2015kta,Boehmer:2015sha,Bettoni:2011fs, Bettoni:2015wla, PhysRevD.95.023515, Koivisto:2015qua, PhysRevD.92.124067, PhysRevD.93.103502, Tamanini:2016klr} for examples of its use in different scenarios)
\begin{equation}\label{eq: Brown Lagrangian}
\mathcal{L}=-\rho\left(n,s\right)+j^{\alpha}\left(\nabla_{\alpha}\phi+s\nabla_{\alpha}\theta+ B_{a}\nabla_{\alpha}A^{a}\right)  \,,  
\end{equation}
where $\rho(s,n)$ is the energy density of the fluid, which depends both on the number density $n$ and on the entropy per particle $s$. Comparing Eq. \eqref{eq: Brown Lagrangian} with the model defined in Eqs. \eqref{eq:Action PF irrotational isentropic} and \eqref{eq:Lagrangian PF irrotational isentropic}, there are additional dynamical variables $s$, $\theta$, $B_{a}$ and $A^{a}$, respectively, with the corresponding equations of motion
\bq\label{eq: equation of motion s}
\frac{\delta S}{\delta s}=&0&=-\frac{\partial\rho}{\partial s}+j^{\alpha}\nabla_{\alpha}\theta\,,  \\
\label{eq: equation of motion theta}
\frac{\delta S}{\delta \theta}=&0&=\nabla_{\alpha}\left(sj^{\alpha}\right)\,,\\
\label{eq: equation of motion B_a}
\frac{\delta S}{\delta A^a}=&0&=j^{\alpha}\nabla_{\alpha}A^{a}\,,\\
\label{eq: equation of motion A^a}
\frac{\delta S}{B_{a}}=&0&=\nabla_{\alpha}\left(j^{\alpha}B_{a}\right)\,.
\eq
These equations, in addition to Eq. \eqref{eq: particle number conservation} and
\begin{equation}\label{eq: 4-velocity potential Brown}
\frac{\delta S}{\delta j^\alpha}=0=\frac{\partial\rho}{\partial n}u_{\alpha}+\nabla_{\alpha}\phi+s\nabla_{\alpha}\theta + B_{a}\nabla_{\alpha}A^{a}\,,
\end{equation}
which replaces Eq. \eqref{eq: 4-velocity potential}, describe the dynamics of the fluid. Here, the scalar field $\theta$ works as a Lagrange multiplier, ensuring that the entropy exchange constraint in Eq. \eqref{eq: equation of motion theta} is satisfied. In combination with the particle number conservation equation [i.e., Eq. \eqref{eq: particle number conservation}] it implies that $j^{\alpha}\nabla_{\alpha}s=0$, which defines an adiabatic flow \cite{1996astro.ph..9119M}. The Lagrange multipliers $B_{a}$ (where $a=1,2,3$) restrict the fluid 4-velocity to be directed along the flow lines of constant $A^{a}$ [Eq. \eqref{eq: equation of motion B_a}], where $A^{a}$ are the Lagrangian coordinates of the fluid. 

The Lagrangian defined in Eq. \eqref{eq: Brown Lagrangian} incorporates some of the most important information for the characterization of a perfect fluid undergoing an adiabatic flow, in the sense that the corresponding dynamical and thermodynamical relations can be elegantly derived from the equations of motion. Despite the extra degrees of freedom present in Eq. \eqref{eq: Brown Lagrangian}, the energy-momentum tensor of a perfect fluid is still recovered with the identifications given in Eqs. \eqref{number n}-\eqref{4velocity} --- even when $\rho$ is a function of both $n$ and $s$. Also, the on-shell Lagrangians given in Secs. \ref{sec: modelII} and \ref{sec: modelIII} can be obtained from Eq. \eqref{eq: Brown Lagrangian}, using Eq. \eqref{eq: particle number conservation} and Eqs. \eqref{eq: equation of motion s}-\eqref{eq: 4-velocity potential Brown}. Although further degrees of freedom can be added (see, e.g., \cite{PhysRevD.22.267,PhysRevD.96.023516}), the Lagrangian presented in Sec. \ref{sec: modelII} (which does not have $s$, $\theta$, $B_{a}$, and $A_{a}$ as dynamical variables) will be sufficient for our discussion of particle conserving isentropic irrotational perfect fluids and their connection with pure $k$-essence scalar field models.

\subsection{Model V}
\label{subsec: model T}
In many situations of interest, a fluid (not necessarily a perfect one) may be simply described as a collection of many identical point particles undergoing quasi-instantaneous scattering from time to time \cite{Avelino2018,Avelino2018a}. Hence, before discussing the Lagrangian of the fluid as a whole, let us start by considering the action of a single point particle with mass $m$
\begin{equation}
S=-\int d \tau \, m \,,
\end{equation}
and energy-momentum tensor 
\begin{equation} \label{eq: particle EM tensor}
T^{*\alpha \beta} = \frac{1}{\sqrt {-g}}\int d \tau \, m \, u^\alpha u^\beta \delta^4\left(x^\mu-\xi^\mu(\tau)\right)\,,
\end{equation}
where the $*$ indicates that the quantity refers to a single particle, $\xi^\mu(\tau)$ represents the particle worldline and $u^\alpha$ are the components of the particle 4-velocity. If one considers its trace $T^*=T^{*\alpha \beta} g_{\alpha \beta}$ and integrates over the whole of spacetime, we obtain
\begin{eqnarray}
\int d^{4}x \sqrt{-g} \, T^* &=&- \int d^4x \,d\tau\, m\, \delta^4\left(x^\mu-\xi^\mu(\tau)\right) \nonumber\\
&=&- \int d\tau \,m \, ,
\end{eqnarray}
which can be immediately identified as the action for a single massive particle, and therefore implies that the corresponding Lagrangian is simply given by
\begin{equation}
\mathcal{L}^*_{\rm on-shell}=T^*\,.
\end{equation}

If a fluid can be modeled as a collection of point particles, then its on-shell Lagrangian at each point will be the average value of the single-particle Lagrangian over a small macroscopic volume around that point
\begin{eqnarray}
\langle \mathcal{L}^*_{\rm on-shell} \rangle &=& \frac{\int d^4 x \sqrt{-g} \, \mathcal{L}^*_{\rm on-shell}}{\int  d^4 x \sqrt{-g}}\\
&=& \frac{\int d^4 x \sqrt{-g} \, T^*}{\int  d^4 x \sqrt{-g}} = \langle T^* \rangle\,,
\end{eqnarray}
where $\langle T^* \rangle = T$ is now the trace of the energy momentum of the perfect fluid. This provides a further possibility for the on-shell Lagrangian of a perfect fluid:
\begin{equation}
\mathcal{L}_{\rm on-shell}= T =-\rho+3p\,,
\end{equation}
where $p=\rho \langle v^2 \rangle/3=\rho \mathcal T$, $\sqrt{\langle v^2 \rangle}$ is the root-mean-square velocity of the particles and $\mathcal T$ is the temperature. Notice that only in the case of dust ($p=0$) do we recover the result obtained for model III ($\mathcal{L}_{\rm on-shell}=-\rho$).

\subsection{Which Lagrangian?}
\label{subsec: which}
We have shown that models I, II, III, IV and V, characterized by different Lagrangians, may be used to describe the dynamics of a perfect fluid. If the matter fields couple only minimally to gravity, then these models may even be used to describe the same physics. However, this degeneracy is generally broken in the presence of NMC either to gravity \cite{Bertolami:2007gv,Bertolami:2008ab,Sotiriou:2008it,Faraoni:2009rk,Harko:2010zi,Bertolami:2010cw,Ribeiro2014,Azizi:2014qsa,Bertolami2014} or to other fields \cite{1982PhRvD..25.1527B,Sandvik:2001rv,Anchordoqui:2003ij,Copeland:2003cv,Lee:2004vm,Koivisto:2005nr,Avelino:2008dc,Ayata:2012,Pourtsidou:2013nha,Boehmer:2015kta,Boehmer:2015sha,Barros:2019rdv,Kase:2019veo}, in which case the identification of the appropriate on-shell Lagrangian may become essential in order to characterize the overall dynamics \cite{Azevedo2018a,Azevedo2019a} (note that this is not an issue if the form of the off-shell Lagrangian is assumed \textit{a priori}, as in \cite{Boehmer:2015kta,Boehmer:2015sha,Bettoni:2011fs, Bettoni:2015wla, PhysRevD.95.023515, Koivisto:2015qua, PhysRevD.92.124067, PhysRevD.93.103502, Tamanini:2016klr}). Models I, II, III and IV, described in the previous section, imply both the conservation of particle number and entropy. However, both the entropy and the particle number are in general not conserved in a fluid described as a collection of point particles. Hence, model V has degrees of freedom that are not accounted for by models I, II, III, and IV. In model V the pressure depends both on the temperature $\mathcal T$ (or, equivalently, the root-mean-square velocity of the particles) and on the energy density $\rho$, with $p= \rho \mathcal T$, while in models I, II, and III $p$ is a function of the number density alone [$p=p(n,s)$ in the case of model IV]. Still, in model V the equation of state parameter $w=p/\rho$ must be in the interval $[0,1/3]$, which while appropriate to describe a significant fraction of the energy content of the Universe, such as CDM, baryons, photons, and neutrinos, cannot be used to describe DE. On the other hand, models I, II, III, IV are specially suited for DE, both because they allow for values of $w \sim -1$ and also because the requirement that $X>0$ can be met only if the spatial variations of the scalar field $\phi$ are sufficiently small. In Sec. \ref{sec:Model} we shall use model II to describe both DE and UDE. However, one should bear in mind that any successful UDE model must account for the observed large scale structure of the Universe, and that a scalar field description of UDE in terms of a perfect fluid is expected to break down on small nonlinear scales \cite{2013PhLB..727...27D}.

\section{\label{sec:NMC} The role of the Lagrangian in NMC models}

As discussed in Sec. \ref{subsec: which}, the energy-momentum tensor does not provide a complete characterization of nonminimally coupled matter fields, since the Lagrangian will also in general explicitly appear in the equations of motion. To further clarify this point, we present a few examples of models in which there is a NMC between matter or radiation with DE or gravity.

\subsection{NMC between matter and DE}

Consider the model described by the following action:
\begin{equation}
    \label{eq: NMC DE}
    S = \int d^4 x \sqrt{-g}\left[R+\mathcal{L}+\mathcal{L}_{\text{F}\phi}\right]\,,
\end{equation}
where $R$ is the Ricci scalar, $\phi$ is the DE scalar field described by the Lagrangian
\begin{equation}
    \label{eq: L DE}
    \mathcal{L} = X - V(\phi)\,,
\end{equation}
and $\mathcal{L}_{\text{F}\phi}$ is the Lagrangian of the matter term featuring a NMC with DE \cite{1995A&A...301..321W,2000PhRvD..62d3511A,2001PhLB..521..133Z,2004ApJ...604....1F}
\begin{equation}
    \label{eq: matter DE nmc}
    \mathcal{L}_{\text{F}\phi}=f(\phi)\mathcal{L}_\text{F}\,.
\end{equation}
Here, $f(\phi)>0$ is a regular function of $\phi$ and $\mathcal{L}_\text{F}$ is the Lagrangian that would describe the matter component in the absence of a NMC to gravity (in which case $f$ would be equal to unity). Using the variational principle it is straightforward to derive the equations of motion for the gravitational and scalar fields. They are given, respectively, by
\begin{equation}
    \label{eq: eom metric NMC DE}
    G^{\alpha\beta} = f\, T_\text{F}^{\,\alpha\beta} +\nabla^\alpha\phi\nabla^\beta\phi -\frac{1}{2}g^{\alpha\beta}\nabla_\mu\phi\nabla^\mu\phi -g^{\alpha\beta}\, V\,,
\end{equation}
\begin{equation}
    \label{eq: eom DE NMC}
    \square\phi -\frac{d V}{d \phi} + \frac{d f}{d \phi}\mathcal{L}_\text{F}=0\,,
\end{equation}
where $G^{\alpha\beta}$ is the Einstein tensor, $\square \equiv \nabla_\alpha\nabla^\alpha$ is the Laplace-Beltrami operator, and 
\begin{equation}
    \label{eq: Em tensor NMC}
    T_\text{F}^{\alpha\beta}=\frac{2}{\sqrt{-g}}\frac{\delta\left(\sqrt{-g}\mathcal{L}_\text{F}\right)}{\delta g_{\alpha\beta}}
\end{equation}
are the components of the energy-momentum tensor associated with the Lagrangian $\mathcal{L}_\text{F}$. Note that the Lagrangian is featured explicitly in the equation of motion for $\phi$. Thus, knowledge of the energy-momentum tensor alone is not enough to fully describe the dynamics of any of the fields.

Consider the coupled matter energy-momentum tensor defined by $T_{\text{F}\phi}^{\alpha\beta}=f(\phi)T_\text{F}^{\alpha\beta}$. By taking the covariant derivative of Eq. \eqref{eq: eom metric NMC DE} and using the Bianchi identities one obtains
\begin{equation}
    \label{eq: EM tensor cons 1}
    \nabla_\alpha T_{\text{F}\phi}^{\alpha\beta} = -\nabla_\beta(\partial^\beta\phi \partial^\alpha\phi) + \frac{1}{2}\nabla^\beta(\partial_\mu\phi \partial^\mu\phi) + \frac{d V}{d\phi}\partial^\alpha\phi \,,
\end{equation}
thus showing that the coupled matter energy-momentum tensor is in general not conserved. Using  Eq. \eqref{eq: eom DE NMC} it is possible to rewrite this equation in such a way as to highlight the explicit dependence on the Lagrangian
\begin{equation}
    \nabla_\beta T_{\text{F}\phi}^{\alpha\beta} = \frac{d f}{d\phi}\mathcal{L}_\text{F}\partial^\alpha\phi \,. \label{EMCF1}
\end{equation}
If $\mathcal{L}_\text{F}$ describes a fluid of particles with fixed rest mass $m_\text{F}$, then one must have $\mathcal{L}_\text{F}=T_\text{F}$, as per Sec. \ref{subsec: model T}. Also, $\mathcal{L}_{\text{F}\phi} =f(\phi)\mathcal{L}_\text{F}$ will describe a fluid with particles of variable rest mass $m(\phi) = f(\phi)m_\text{F}$. In this case, Eq.~\eqref{EMCF1} may also be written as
\begin{equation}
    \nabla_\mu T_{\text{F}\phi}^{\alpha\mu} = -\beta T_\text{F}\partial^\alpha\phi \,, \label{EMCF2}
\end{equation}
where
\begin{equation}
    \beta(\phi) = -\frac{d \ln m(\phi)}{d \phi}\,. \label{betadef}
\end{equation}
In the present paper we shall focus on the macroscopic fluid dynamics, but the NMC between matter and DE also affects the dynamics of the individual particles (see, for example, \cite{Ayata:2012} for more details).

\subsubsection{Coupling between DE and neutrinos}

A related model featuring a NMC between neutrinos and DE, so-called growing neutrino quintessence, where the neutrinos are described the Lagrangian
\begin{equation}
    \label{eq: neutrinos L}
    \mathcal{L}_{\mathcal V} = i\bar{\psi}\left(\gamma^\alpha\nabla_\alpha +m(\phi)\right)\psi  \,,
\end{equation}
has been investigated in \cite{Ayata:2012}. Here, $\bar{\psi}$ is the Dirac conjugate, $m (\phi)$ is a DE-field dependent neutrino rest mass, the quantities $\gamma^\alpha(x)$ are related to the usual Dirac matrices $\gamma^a$ via $\gamma^\alpha =\gamma^a e^\alpha_a$ where $e^\alpha_a$ are the vierbein, with $g^{\alpha\beta}=e_a^\alpha e_b^\beta \eta^{ab}$ and $\eta^{ab} =\text{diag}(-1,1,1,1)$, and $\nabla_\alpha$ is the covariant derivative that now takes into account the spin connection (see \cite{RevModPhys.29.465} for more details on the vierbein formalism). The classical equations of motion for the neutrinos, derived from the action
\begin{equation}
    \label{eq: neutrino action}
    S = \int d^4 x \sqrt{-g}\left[R+\mathcal{L}+\mathcal{L}_{\mathcal V}\right]\,,
\end{equation}
may be written as
\begin{align}
    \label{eq: neutrino eom}
    \gamma^\alpha\nabla_\alpha \psi+m(\phi)\psi &=0\,, \\
    \nabla_\alpha\bar{\psi}\gamma^\alpha-m(\phi)\bar{\psi} &= 0\,.
\end{align}
The components of the corresponding energy-momentum tensor are \cite{Ayata:2012}
\begin{equation}
    \label{eq: neutrino em tensor}
    T_{\mathcal V}^{\alpha\beta} = -\frac{i}{2}\bar{\psi}\gamma^{(\beta}\nabla^{\alpha)}\psi +\frac{i}{2}\nabla^{(\alpha}\bar{\psi}\gamma^{\beta)}\psi \,,
\end{equation}
where the parentheses represent a symmetrization over the indices $\alpha$ and $\beta$. The trace of the energy-momentum tensor is given by \cite{Ayata:2012}
\begin{equation}
    \label{eq: EM neutrino trace}
    T_{\mathcal V} = i\bar\psi\psi m(\phi) = - m(\phi)  \widehat n \,,
\end{equation}
where $\widehat n= -i\bar\psi\psi$ is a scalar that in the nonrelativistic limit corresponds to the neutrino number density.

Taking the covariant derivative of Eq. \eqref{eq: neutrino em tensor} one obtains
\begin{equation}
    \label{eq: EM neut tensor cons1}
    \nabla_\mu T_{\mathcal V}^{\alpha\mu} = -\beta (\phi) T_{\mathcal V}\partial^\alpha\phi \,,
\end{equation}
where $\beta(\phi)$ is defined in Eq. \eqref{betadef}. A comparison between Eqs. \eqref{EMCF2} and \eqref{eq: EM neut tensor cons1} implies that $\mathcal{L}_{F\phi}$ and $\mathcal{L}_{\mathcal V}$ provide equivalent on-shell descriptions of a fluid of neutrinos in the presence of a NMC to gravity. The same result could be achieved by analyzing the dynamics of individual neutrino particles \cite{Ayata:2012}. 

\subsubsection{Coupling between DE and the electromagnetic fields}

Consider now a model described by Eqs. \eqref{eq: NMC DE} and \eqref{eq: matter DE nmc} with
\begin{equation}
    \label{eq: electro lagrangian}
    \mathcal{L}_\text{F} = \mathcal{L}_\text{EM} = -\frac{1}{4}F_{\alpha\beta}F^{\alpha\beta} \,,
\end{equation}
where $F_{\alpha\beta}$ is the electromagnetic field tensor \cite{1982PhRvD..25.1527B,Sandvik:2001rv,Avelino:2008dc}. This model will naturally lead to a varying fine-structure ``constant"
\begin{equation}
    \label{eq: fine struct}
    \alpha(\phi) = \frac{\alpha_0}{f(\phi)} \,,
\end{equation}
whose evolution is driven by the dynamics of the DE scalar field $\phi$. Equation \eqref{eq: eom DE NMC} implies that the corresponding equation of motion is given by 
\begin{equation}
    \label{eq: eom electro scalar}
    \square\phi -\frac{d V}{d \phi} + \frac{\alpha_0}{4
    \alpha^2}\frac{d \alpha}{d \phi}F_{\alpha\beta}F^{\alpha\beta}= 0 
\end{equation}
or, equivalently,
\begin{equation}
    \label{eq: eom electro scalar 2}
    \square\phi -\frac{d V}{d \phi} - \frac{\alpha_0}{    \alpha^2}\frac{d \alpha}{d \phi}\mathcal{L}_\text{EM}= 0 \,.
\end{equation}
Electromagnetic contributions to baryon and lepton mass mean that in general $\mathcal{L}_\text{EM} \neq 0$. However, $\mathcal{L}_\text{photons}=(E^2-B^2)_\text{photons} = 0$ (here, $E$ and $B$ represent the magnitude of the electric and magnetic fields, respectively) and, therefore, electromagnetic radiation does contribute to $\mathcal{L}_\text{EM}$. Note that the last term on the left-hand side of Eq. \eqref{eq: eom electro scalar} is constrained, via the equivalence principle, to be small \cite{Olive:2001vz}. Therefore, the contribution of this term to the dynamics of the DE field is often disregarded (see, e.g., \cite{Anchordoqui:2003ij,Copeland:2003cv,Lee:2004vm}).

It is common, in particular in cosmology, to describe a background of electromagnetic radiation as a fluid of point particles whose rest mass is equal to zero (photons). In this case one should use the appropriate on-shell Lagrangian of this fluid in Eq. \eqref{eq: eom electro scalar 2}. In Sec. \ref{sec:Perfect fluid lagrangian} we have shown that if the fluid is made of particles of fixed mass, then the appropriate on-shell Lagrangian is $\mathcal{L}_\text{EM}= T = 3p-\rho$. For photons (with $p=\rho/3$) this again implies that the on-shell Lagrangian $\mathcal{L}_\text{EM}$ vanishes, thus confirming that photons do not source the evolution of the DE scalar field $\phi$. 

\subsection{NMC between matter and gravitational fields}

A different type of NMC occurs in theories that feature a direct coupling between a function of the Ricci scalar and the Lagrangian of the matter fields \cite{Bertolami:2007gv,Bertolami:2008ab,Sotiriou:2008it,Faraoni:2009rk,Harko:2010zi}. The simplest of these models is described by the Lagrangian
\begin{equation}
\label{eq:action}
S=\int d^4 x \sqrt{-g} \left[R+ f(R)\mathcal{L}_\text{m}\right]\,.
\end{equation}
The corresponding equations of motion for the gravitational field are given by 
\begin{align}
	\label{eq:eqmotion}
	\left(1  + f'\mathcal{L}_\text{m}\right) G^{\alpha\beta} =& \frac{1}{2}  f \, T^{\alpha\beta} +\Delta^{\alpha\beta}\left(f'\mathcal{L}_\text{m}\right) \nonumber \\
	&-\frac{1}{2} R f' \mathcal{L}_\text{m} g^{\alpha\beta}\, , 
\end{align}
where a prime denotes a derivative with respect to $R$, $\Delta^{\alpha\beta} \equiv \nabla^\alpha \nabla^\beta - g^{\alpha\beta} \Box$, and 
\begin{equation}
    \label{eq: Em tensor NMC1}
    T^{\alpha\beta}=\frac{2}{\sqrt{-g}}\frac{\delta\left(\sqrt{-g}\mathcal{L}_\text{m}\right)}{\delta g_{\alpha\beta}}
\end{equation}
are the components of the energy-momentum tensor. The covariant derivative of Eq. \eqref{eq:eqmotion} gives
\begin{equation}
\label{eq:noncons}
\nabla_\beta T^{\alpha\beta}=(g^{\alpha\beta}\mathcal{L}_m-T^{\alpha\beta})\nabla_\beta \ln f\,,
\end{equation}
where the explicit dependence on the Lagrangian is once again evident --- notice that due to the NMC to gravity the energy-momentum tensor is no longer conserved. Moreover, since the matter fields are nonminimally coupled to the geometry, an additional acceleration term should be added to the geodesic equation
\begin{equation}
\label{eq:geodesicf}
\frac{du^{\alpha}}{d\tau}+\Gamma^\alpha_{\mu\nu} u^\mu u^\nu = \mathfrak{a}^\alpha \,,
\end{equation}
which, in the case of a perfect fluid, can be written as
\begin{equation}
\label{eq:accelerationf}
\mathfrak{a}^\alpha=\frac{1}{\rho+p}\left[(\mathcal{L}_\text{m}-p)\nabla_\beta \ln f - \nabla_\beta p\right]h^{\alpha\beta}\,,
\end{equation}
where $h^{\mu \nu}=g^{\mu \nu}+ u^\mu u^\nu$ is the projection operator. The use of the appropriate Lagrangian is then crucial when constraining these theories, in particular using cosmic microwave background or big-bang nucleosynthesis observations \cite{Avelino2018, Azevedo2018a} (see also \cite{Azevedo2019a,Avelino:2020fek}). \\

As demonstrated in Sec. \ref{sec:Perfect fluid lagrangian}, and illustrated in the previous examples (NMC between neutrinos or photons and DE), the condition $\mathcal L_m=T$ needs to be satisfied in any equivalent on-shell fluid description of models featuring point particles of fixed mass. This condition, however, does not generally hold in the case of DE or UDE.


\section{\label{sec:Model} Mapping DE into UDE}

The main feature of most UDE models is that of mimicking DE and CDM with a single underlying perfect fluid or scalar field (see \cite{Avelino:2007tu} for a discussion of the single fluid hypothesis). To construct a model with these properties we shall consider the Lagrangian 
\begin{equation}\label{eq: Widetilde Lagrangian}
\mathcal{L}_{\rm ude}=\mathcal{L}_{\rm de} + \mathcal{L}_{\rm m}\,.
\end{equation}
Here, we shall assume that $\mathcal{L}_{\rm de} \equiv \mathcal{L}_{\rm de}\left(X\right)$ is an arbitrary pure kinetic DE Lagrangian and that the ratio between $\mathcal{L}_{\rm m}$ and $\mathcal{L}_{\rm de}$ vanishes on-shell (or is extremely small, so that the contribution of $\mathcal{L}_{\rm m}$ to the total pressure can be neglected). Therefore, the UDE Lagrangian $\mathcal{L}_{\rm ude}$ describes a fluid with proper pressure $p_{\rm ude}=\mathcal{L}_{\rm ude(on-shell)}=\mathcal{L}_{\rm de(on-shell)}=p_{\rm de}$ and energy density
\begin{equation}
\rho_{\rm ude}=\rho_{\rm de}+\rho_{\rm m}\,,\label{eq: ude density}
\end{equation}
where $\rho_{\rm de}=2X\mathcal{L}_{{\rm de},X}-\mathcal{L}_{\rm de}$. The new Lagrangian may be regarded as a UDE model provided that $w_{\rm de}=p_{\rm de}/\rho_{\rm de} \sim -1$ or, equivalently, $\rho_{\rm de}={\mathcal L}_{\rm de(on-shell)}/ w_{\rm de} \sim -{\mathcal L}_{\rm de(on-shell)}$.

\subsection{\texorpdfstring{$\Lambda$}{L}CDM as a UDE model}

One possible choice for $\mathcal{L}_{\rm cdm}$ would be to consider 
\begin{equation}\label{eq: cdm Lagrange multiplier}
\mathcal{L}_{\rm cdm}=\lambda\left(X-V\left(\phi\right)\right)\,,
\end{equation}
where $\lambda$ is a Lagrange multiplier and $V\left(\phi\right)>0$ is a function of $\phi$ \cite{2011PhLB..702..107G,2010JCAP...05..012L}. This choice ensures that the constraint $X=V(\phi)$ is always satisfied on-shell, thus implying that $\mathcal{L}_{\rm cdm(on-shell)}=0$ or, equivalently, that $p_{\rm ude}=\mathcal{L}_{\rm ude(on-shell)}=\mathcal{L}_{\rm de(on-shell)}=p_{\rm de}$. On the other hand, the density of the UDE fluid is given by Eq. \eqref{eq: ude density} with 
\begin{equation}
\rho_{\rm cdm}=\lambda\left(X+V\left(\phi\right)\right)=2 \lambda X \,.
\end{equation}
Note that the Lagrange multiplier $\lambda$ is a dynamical field whose evolution is such as to ensure that the energy-momentum tensor of the UDE fluid, subject to the constraint $X=V(\phi)$, is covariantly conserved. In the particular case with $V(\phi)=V_0=\rm const$ one would get $X=V_0=\rm const$, thus implying that $p_{\rm ude}=\mathcal{L}_{\rm de(on-shell)}$ would be a constant. Hence, such a UDE model would be totally equivalent to $\Lambda \rm CDM$ \cite{Avelino:2003cf,Avelino:2007tu}. In general, however, $\rho_{\rm ude}$ is a function of $X$ and $\lambda$, where both are dynamical variables. Hence, these models do not generally belong to the class of irrotational perfect fluid models with the conserved particle number and constant entropy per particle considered in Sec. \ref{sec: modelII} which have $\rho=\rho(X)$ and $p=p(X)$. \\

An alternative would be to consider a class of purely kinetic Lagrangians given by \cite{Mukhanov:2005sc,Avelino:2008cu}
\begin{equation}\label{eq: k essence X to the power of gamma}
\mathcal{L}\left(X\right)= AX^{\gamma}\,,
\end{equation}
where $A$ and $\gamma$ are positive real constants. These models describe an isentropic perfect fluid with pressure $p=\mathcal{L}(X)$ and energy density
\begin{equation}
\rho=2X\mathcal{L}_{,X}-\mathcal{L}=\left(2\gamma-1\right)AX^{\gamma}\,,
\end{equation}
with the equation of state parameter
\begin{equation}
w \equiv \frac{p}{\rho} = \frac{1}{2\gamma-1}\,,
\end{equation}
being a constant. In the $\gamma\rightarrow\infty$ limit $w\rightarrow 0$. Hence, this fluid mimicks pressureless dust in this limit. Thus another possible choice for $\mathcal{L}_{\rm m}$ would be
\begin{equation}
\mathcal{L}_{\rm m}\left(X\right)=\lim_{\gamma\rightarrow\infty} A(\gamma) X^{\gamma}\,.
\end{equation}
The function $A(\gamma)$ is chosen in such a way that $p_{\rm m}$ vanishes at every spacetime point in this limit, but 
\begin{equation}
\rho_{\rm m}=\lim_{\gamma\rightarrow\infty}\left(2\gamma-1\right)A(\gamma)X^{\gamma}
\end{equation}
is essentially unrestricted. Note that by choosing $A(\gamma)$ such that the function $C(\gamma)=(2\gamma-1)A(\gamma)$ tends to a constant $C_\infty$ in the $\gamma \to \infty$ limit, $X$ must be equal to unity in this limit. Note, however, that the density may take any value in this limit since $1^\infty$ is indeterminate. On the other hand, in the $\gamma \to \infty$ limit the equation of motion of the scalar field 
\be
\left(\mathcal L_{,X}g^{\alpha \beta} + \mathcal L_{,XX} \nabla^\alpha \phi\nabla^\beta \phi \right) \nabla_\alpha \nabla_\beta \phi=0\,,
\ee
reduces to
\be
\nabla^\alpha \phi\nabla^\beta \phi  \nabla_\alpha \nabla_\beta \phi= -\nabla^\alpha \phi \nabla_\alpha X =  0\,,
\ee
thus implying that the equation of motion does indeed preserve the condition $X=1$ in this limit. Also, note that since the condition $X>0$ is always satisfied this model describes the dynamics of a perfect fluid. The caveat is that the corresponding UDE model would have $p_{\rm ude}=\mathcal{L}_{\rm de(on-shell)}\left(X=1\right)=\rm const$ and, therefore, would again be totally equivalent to $\Lambda \rm CDM$.

\subsection{Mapping $k$-essence models with the same on-shell Lagrangian}

Consider an isentropic perfect fluid with proper pressure and density $p=p(\mu)$, $\rho=\rho(\mu)$ (with $\mu=\mu(n)$), and 4-velocity $\bf u$ at each spacetime point. The transformation
\begin{eqnarray}
\label{eq: rescaling rho}
\widetilde{\rho} &=&\rho+mn \,,\\
\label{eq: rescaling mu}
\widetilde{\mu}&=&\mu+m \,,
\end{eqnarray}
at every point with the 4-velocity unchanged leads to a different perfect fluid, but leaves the proper pressure unaltered, so that ${\widetilde p}\left(\widetilde{\mu}\right)=p\left(\mu\right)$ (here, $m>0$ is a constant) --- i.e. the transformations given in Eqs. \eqref{eq: rescaling rho} and \eqref{eq: rescaling mu} leave Eq. \eqref{eq: 1st law legendre transformation} invariant. Note that, if the original fluid represented a constant density with $p=-\rho={\rm const}$ (a cosmological constant), then this transformation would simply add a pressureless dustlike component to the original DE fluid.

Consider the case in which one starts with a perfect fluid described by a purely kinetic Lagrangian $\mathcal{L}(X)=p(X)$, with $\mu^2=2X$. Let us also write the Lagrangian of the new fluid as $\mathcal{\widetilde L}(\widetilde X)=\mathcal{L}(X)$ and its 4-velocity as $\widetilde u^{\alpha} = -\nabla^{\alpha}\widetilde\phi /\sqrt{2\widetilde X}$, where $\widetilde X = - \nabla_\alpha {\widetilde \phi} \nabla^\alpha {\widetilde \phi}/2$ and ${\widetilde \mu}^2 = 2 \widetilde X$. Writing Eq. \eqref{eq: rescaling mu} as $\sqrt{2\widetilde{X}}=\sqrt{2X}+m$ we get the following relation between the kinetic terms $X$ and $\widetilde X$:
\begin{equation}\label{eq:XY relation}
{\widetilde X}=X+m\sqrt{2X}+\frac{m^2}{2}\,.
\end{equation}

The energy-momentum tensor of the new fluid may be written as
\begin{eqnarray}\label{eq: EMT with matter term}
{\widetilde T}^{\alpha\beta}&=&\mathcal{\widetilde L}_{,{\widetilde X}}\nabla^{\alpha}{\widetilde \phi}\nabla^{\beta}{\widetilde \phi}+\mathcal{\widetilde L}g^{\alpha\beta}\nonumber=2 {\widetilde X} \mathcal{\widetilde L}_{,{\widetilde X}}{\widetilde u}^{\alpha}{\widetilde u}^{\beta}+\mathcal{\widetilde L}g^{\alpha\beta}\nonumber\\
&=&\sqrt{2X}\left(\sqrt{2X}+m\right)\mathcal{L}_{,X}u^{\alpha}u^{\beta}+\mathcal{L} g^{\alpha\beta}\,,\nonumber\\
&=& T^{\alpha\beta}+ T_m^{\alpha\beta}\,,
\end{eqnarray}
where $T^{\alpha\beta}$ is the energy-momentum tensor of the original fluid given in Eq. \eqref{eq: EMT k-essence X} and
\begin{equation}
T^{\alpha\beta}_{m}=\rho_{\rm m}u^{\alpha}u^{\beta} \qquad \rho_{\rm m}=mn=m\sqrt{2X}\mathcal{L}_{,X}\,,
\end{equation} 
is an additional dustlike component.
Here, we have used Eq. \eqref{eq:XY relation} and the relations 
\begin{eqnarray}\label{eq: X-X tilde relation}
X_{,{\widetilde X}}&=&\left({\widetilde X}_{,X}\right)^{-1}=\frac{\sqrt{2X}}{\sqrt{2X}+m}\,,\\
u^{\alpha}&=&-\frac{\nabla^{\alpha}\phi}{\sqrt{2X}}=-\frac{\nabla^{\alpha}{\widetilde\phi}}{\sqrt{2{\widetilde X}}}={\widetilde u}^{\alpha}\,. \label{4-vel}
\end{eqnarray} 
Equation \eqref{4-vel} is equivalent to
\begin{equation}
\nabla^{\alpha}{\widetilde \phi}=\frac{\sqrt{2X}+m}{\sqrt{2X}}\nabla^{\alpha}\phi\,, \label{phis}
\end{equation}
but, unfortunately, given a scalar field $\phi$ it may not always be possible to find another scalar field $\widetilde \phi$ which satisfies this equation. However, in a perfectly homogeneous and isotropic Friedmann-Lema\^itre-Robertson-Walker (FLRW) universe $\phi$ and $X$ are functions of cosmic time alone, making it possible to define a scalar field $\widetilde \phi$ fulfilling Eq. \eqref{phis}. Given that the DE field is expected to be nearly homogeneous this will turn out to be the most relevant case, which will be further explored later in the paper. Also, having defined $\mathcal{\widetilde L}(\widetilde X)$ it is possible to explore the full consequences of the model, taking into account cosmological perturbations.

The energy momentum tensor is covariantly conserved or equivalently, $\nabla_\alpha {\widetilde T}^{\alpha \beta}=0$. This implies that 
\begin{eqnarray}
\label{eq: Coupled k essence DE}
\nabla_\alpha T^{\alpha\beta}&=&Q^\beta\,,\\
\label{eq: Coupled k essence DM}
\nabla_\alpha T_m^{\alpha\beta}&=&-Q^\beta\,,
\end{eqnarray}
where $Q^{\beta}$ is the coupling between the two components.
Contracting the equation $\nabla_\alpha {\widetilde T}^{\alpha \beta}=0$ with ${\widetilde u}_\beta$, one obtains the continuity equation

\begin{equation}
\widetilde{u}^{\alpha}\nabla_{\alpha}\widetilde{\rho}+\left(\widetilde{\rho}+\widetilde{p}\right)\nabla_{\alpha}\widetilde{u}^{\alpha}=0\,,    
\end{equation}
which is equivalent to the equation of conservation of the particle number $\nabla_{\alpha}\left(\mathcal{\widetilde L}_{,{\widetilde X}} \sqrt {2 {\widetilde X}} {\widetilde u}^\alpha \right) =0$ [see Eq. \eqref{eq: k-essence eq. of motion}]. Given that $n=\sqrt{2X}{\mathcal{L}}_{,X}=\sqrt{2 {\widetilde X}}{\mathcal{\widetilde L}}_{,{\widetilde X}}={\widetilde n}$ and ${\widetilde u}_\beta=u_\beta$, the particle number conservation equation may also be written as $\nabla_{\alpha}\left(\mathcal{L}_{,X} \sqrt {2  X} u^\alpha \right)=0$. Taking this into account, it is simple to show that $Q^\beta u_\beta=0$. 
The contraction of $\nabla_\alpha {\widetilde T}^{\alpha \beta}=0$ with $h^{\nu}_{\beta}=\delta^{\nu}_{\beta} + u^{\nu}u_{\beta}$ (where $\delta^{\nu}_{\beta}=g^{\nu\alpha}g_{\alpha\beta}$ is the Kronecker delta) results in
\begin{equation}
\left(g^{\nu\alpha}+\widetilde{u}^{\nu}\widetilde{u}^{\alpha}\right)\nabla_{\alpha}\widetilde{p}=-\left(\widetilde{\rho}+\widetilde{p}\right)\widetilde{a}^{\nu}\,,    
\end{equation}
with $\widetilde{a}^{\alpha}=\widetilde{u}^{\beta}\nabla_{\beta}\widetilde{u}^{\alpha}=a^{\alpha}$ being the components of the 4-acceleration (notice that $\widetilde{u}_{\alpha}\widetilde{a}^{\alpha}=0$). From the contraction of $h^{\nu}_{\beta}$ with Eq. \eqref{eq: Coupled k essence DM} one finds that
\begin{equation}
\rho_{\rm m}a^{\nu}=-h^{\nu}_{\beta}Q^{\beta}\,. \label{4-acc}
\end{equation}
\\
In a perfectly homogeneous and isotropic FLRW background $u^0=1$ and $a^\nu=0$. Hence, Eq. \eqref{4-acc} in combination with the condition $Q^\beta u_\beta=0$, implies that $Q^\nu=0$. In this case, the energy-momentum tensors of the matter and DE components are separately conserved.

\subsubsection{Background evolution}

In a FLRW homogeneous and isotropic universe Eq. \eqref{eq: k-essence eq. of motion} has the known solution \cite{2004PhRvL..93a1301S,2004PhRvD..69l3517C}
\begin{equation}\label{eq: k-essence FRW solution}
X\mathcal{L}_{,X}^{2} \propto a^{-6} \propto (1+z)^6\,,
\end{equation}
where $a$ is the scale factor and $z \equiv 1/a-1$ is the redshift (the scale factor $a$ is normalized to unity at the present time). Therefore, we may write $n=\sqrt{2X} \mathcal {L}_{,X} =n_0 (1+z)^3$, where $n_0 \equiv n(z=0)$. Hence, $\rho_{\rm m} = m n = m n_0  (1+z)^3$, irrespective of the original pure $k$-essence model. Thus, the equation of state parameter of the transformed fluid is given by
\begin{equation}\label{eq: w UDE model}
{\widetilde w}\equiv\frac{\widetilde p}{\widetilde \rho}=\frac{p}{\rho+mn}=\frac{w}{1+mn_0 (1+z)^3/{\rho}} \,,
\end{equation}
where $w \equiv p/\rho$ is the equation of state parameter of the original fluid. On the other hand, the sound speed of the transformed fluid, defined by ${\widetilde c}_s^2 \equiv{{\widetilde p}_{,\widetilde X}}/{{\widetilde \rho}_{,\widetilde X}}$  \cite{1999PhLB..458..219G} is equal to
\begin{equation}\label{eq: k-essence FRW solution 2}
{\widetilde c}_s^2 = \frac{\widetilde p_{,z}}{\widetilde \rho_{,z}} = \frac{c_s^2}{1+3 mn_0 (1+z)^2/\rho_{,z}}  \,,
\end{equation}
where $c_s^2 \equiv p_{,X}/\rho_{,X}$ is the sound speed of the original fluid. Hence, given $m$ and $n_0$, the evolution of the sound speed squared of the transformed fluid ${\widetilde c}_s^2$ with the redshift is completely determined by the evolution of the sound speed squared $c_s^2$ and of the density of the original model.

\subsubsection{Cosmological perturbations}

In this subsection we shall briefly consider the linear evolution of metric and density perturbation in these models (see, e.g., \cite{Bartolo:2003ad,Bamba:2011ih}). In the longitudinal gauge the line element may be written as
\begin{equation}
ds^{2}=-\left(1+2\Phi\right)dt^{2}+a^{2}\left(t\right)\left(1-2\Psi\right)\delta_{ij}dx^{i}dx^{j}\,.
\end{equation}
In the case of a perfect fluid the anisotropic stress vanishes, thus implying that $\Phi=\Psi$ \cite{Mukhanov:2005sc} (here, $\Phi$ is the Newtonian gravitational potential). 

Let us write $\phi=\phi_{(\rm{b})}+\delta\phi$, where the subscript ``$(\rm{b})$'' refers to the purely time dependent background value of $\phi$ and $\delta \phi$ denotes the fluctuation of $\phi$ with respect to the background value (we shall use the same notation in the case of the other variables). At first order in $\delta \phi$, the energy-momentum tensor defined in Eq. \eqref{eq: EMT with matter term} may be written as
\begin{equation}
\widetilde{T}^{\alpha}_{\beta}=\widetilde{T}^{\alpha}_{\beta(\rm{b})}+\delta \widetilde{T}^{\alpha}_{\beta}\,, \label{EM1}
\end{equation}
where $\widetilde{T}^{0}_{0(\rm{b})}=-\widetilde{\rho}_{(\rm{b})}$, $\widetilde{T}^{i}_{j(\rm{b})}=\widetilde{p}_{(\rm{b})} \, \delta^{i}_{j}$,  $\delta^{\alpha}_{\beta}=g^{\alpha\gamma}g_{\beta\gamma}$ is the Kronecker delta, and
\begin{eqnarray}
\delta \widetilde{T}^{0}_{0}&=&-\delta\widetilde{\rho} =- \left(\mathcal{L}_{,X(\rm{b})}+2X_{(\rm{b})}\mathcal{L}_{,XX(\rm{b})}\right)\nonumber\\ &\times&\left(1+\frac{m}{\sqrt{2X_{(\rm{b})}}}\right)\delta X\,,\label{EM2}\\
\delta\widetilde{T}^{0}_{i}&=&\mathcal{L}_{,X(\rm{b})}\left(\dot{\phi}_{(\rm{b})}+m\right)\dot{\phi}_{(\rm{b})}\delta u_{i}\,,\label{EM3}\\
\delta\widetilde{T}^{i}_{0}&=&-a^{-2}\mathcal{L}_{,X(\rm{b})}\left(\dot{\phi}_{(\rm{b})}+m\right)\dot{\phi}_{(\rm{b})}\delta u_{i}\,,\\
\delta\widetilde{T}^{i}_{j}&=&\delta\widetilde{p}\,\delta^{i}_{j}=\mathcal{L}_{,X(\rm{b})}\delta^{i}_{j} \delta X\,.\label{EM4}
\end{eqnarray}
Here, a dot denotes a derivative with respect to physical time, $\delta u_{i}=a^{2}\delta u^{i}$, and it has been taken into account that, up to first order in $\delta \phi$, the perturbation to the kinetic term $X$ is given by
\begin{equation}
\delta X= 2X_{(\rm{b})}\left(\frac{\dot{\delta\phi}}{\dot{\phi}_{(\rm{b})}}-\Phi\right)\,.
\end{equation}

Given the energy-momentum tensor defined by Eqs. \eqref{EM1}-\eqref{EM4}, the Einstein equations imply that, up to first order in $\delta \phi$ and $\Phi$,
\begin{eqnarray}
\label{eq: Einstein equation perturbation 00}
-\frac{\Delta\Phi}{a^{2}}+3H\left(\dot{\Phi}+H\Phi\right)&=&4\pi G\delta \widetilde{T}^{0}_{0}\,,\\
\label{eq: Einstein equation perturbation 0i}
\nabla_{i}\left(\dot{\Phi}+H\Phi\right)&=&4\pi G\delta \widetilde{T}^{0}_{i}\,,\\
\label{eq: Einstein equation perturbation ij}
\left[\ddot{\Phi}+4H\dot{\Phi}+\left(2\dot{H}+3H^{2}\right)\Phi\right]\delta^{i}_{j}&=&4\pi G\delta \widetilde{T}^{i}_{j}\,,
\end{eqnarray}
where $\Delta$ denotes the Laplacian. Equations \eqref{eq: Einstein equation perturbation 00} and \eqref{eq: Einstein equation perturbation ij} may then be combined to obtain the following equation for the evolution of the gravitational potential
\begin{eqnarray}
\label{gravpotev}
\ddot{\Phi}&+&H\dot{\Phi}\left(4+3\widetilde{c}^{2}_{s}\right)+2\dot{H}\Phi+3H^{2}\left(1+\widetilde{c}^{2}_{s}\right)\Phi \nonumber\\
&=&\widetilde{c}^{2}_{s}\frac{\Delta\Phi}{a^{2}}\,,
\end{eqnarray}
with
\begin{equation}\label{eq: sound speed UDE}
\widetilde{c}_{s}^{2}=\frac{\sqrt{2X_{(\rm{b})}}}{\sqrt{2X_{(\rm{b})}}+m}\frac{\mathcal{L}_{,X(\rm{b})}}{\mathcal{L}_{,X(\rm{b})}+2X_{(\rm{b})}\mathcal{L}_{,XX(\rm{b})}}\,.
\end{equation}
It is straightforward to show that this expression for $\widetilde{c}_{s}^{2}$ is consistent with the one given in Eq. \eqref{eq: k-essence FRW solution 2} and that, for $m \gg X_{(\rm{b})}$ (or, equivalently, $|\rho_{,z}/n_{,z}| \ll m$), one has $\widetilde{c}_{s}^{2} \ll {c}_{s}^{2}$. Also, from the Fourier transform of Eq.  \eqref{gravpotev} one may check that small scale pathological instabilities are avoided as long as  $\widetilde{c}_{s}^{2} \ge 0$.

\subsection{Nontrivial map between DE and UDE models}

In this subsection we shall assume that the original Lagrangian $\mathcal L (X)$ describes a DE fluid with equation of state parameter $w_{0} = w_{\rm de 0} \sim -1$, so that the transformed Lagrangian $\widetilde{\mathcal L} (\widetilde X)$ defines a UDE fluid with equation of state parameter $\widetilde{w} = w_{\rm ude}$ (in the following, we shall use the subscripts ``$\rm de$'' and ``$\rm ude$'', respectively, when referring to DE and UDE). In this context, the equation of state parameter of the UDE fluid may be written as [see Eq. \eqref{eq: w UDE model}]
\begin{equation}
w_{\rm ude}(z)=\frac{w_{\rm de}(z)}{1+m n_0(1+z)^3/\rho_{\rm de}(z)}\,.
\end{equation}
Since this model is defined by a purely kinetic Lagrangian, the sound speed coincides with the adiabatic sound speed given by
\begin{equation}
{c}_{\rm s(ude)}^{2} = \frac{p_{{\rm ude},z}}{\rho_{{\rm ude},z}} = \left(1+3\frac{m n_0\left(1+z\right)^{2}}{\rho_{{\rm de},z}}\right)^{-1}c_{\rm s(de)}^{2}\,,
\end{equation}
where $c_{\rm s(de)}^{2}=p_{{\rm de},X}/\rho_{{\rm de},X}=p_{{\rm de},z}/\rho_{{\rm de},z}$ is the sound speed of the original DE fluid. Notice that, as long as the sound speed squared $c^2_{\rm s(de)}$ of the input DE fluid is positive, the same is verified in the case of the resulting UDE fluid, thus ensuring that no pathological instabilities occur (at a nonlinear level it is guaranteed \textit{a priori} by the fact that the behavior of UDE is similar to that of CDM in the high density regime). If $\rho_{{\rm de},z}> -3m n_0\left(1+z\right)^{2}$ then $c^2_{\rm s(de)}>0$ is required in order to guarantee that ${c}_{\rm s(ude)}^{2}>0$. On the other hand, if $\rho_{{\rm de},z}<-3m n_0\left(1+z\right)^{2} < 0$ the condition $c^2_{\rm s(ude)}>0$ would be satisfied if, and only if, $c^2_{\rm s(de)} < 0$. However, we shall not explore this case in the present paper, since it would require the consideration of phantom DE models.

\subsubsection{Input DE model: \texorpdfstring{$w_{\rm de}=\rm const$}{wde=const}}

It is instructive to start by examining a DE model with constant $w_{\rm de} \sim -1$ (here, we shall consider a nonphantom DE model with $w_{\rm de}>-1$) defined by the Lagrangian
\be
\mathcal L (X) = C X^\frac{1+w_{\rm de}}{2w_{\rm de}}\,,
\ee
where $C<0$ is a constant (notice that a constant $w_{\rm de}$ implies that $c^2_{\rm s(de)}=w_{\rm de}$). In this case,
\be
\widetilde {\mathcal L} (\widetilde X) = C \left(\sqrt {\widetilde X} - \frac{m}{\sqrt 2 }\right)^\frac{1+w_{\rm de}}{w_{\rm de}}
\ee
describes a UDE model with proper energy density
\bq
\rho_{\rm ude} (\widetilde X) &=& 2 \widetilde X \widetilde {\mathcal L}_{,\widetilde X} - \widetilde {\mathcal L} = C \left(\frac{\sqrt {\widetilde X}}{w_{\rm de}} + \frac{m}{\sqrt 2 }\right)\nonumber\\
&\times&\left(\sqrt {\widetilde X} - \frac{m}{\sqrt 2 }\right)^\frac{1}{w_{\rm de}}
\eq
and proper pressure $p_{\rm ude} (\widetilde X) =  \widetilde {\mathcal L} (\widetilde X)={\mathcal L} (X)$. Notice that $\rho_{\rm ude} \to \infty$ for ${\widetilde X}  \to m^2/2$ (dark matter limit, with $p_{\rm ude} \to 0$), and that
\be
\rho_{\rm ude} \sim \frac{C}{w_{\rm de}} X^{\frac{w_{\rm de}+1}{w_{\rm de}}}  \to  0
\ee
in the  ${\widetilde X} \to \infty$ limit (dark energy limit, satisfying $p_{\rm ude} \sim  w_{\rm de} \rho_{\rm ude} \sim - \rho_{\rm ude} $)  --- for ${\widetilde X} \in \ ]m^2/2,+\infty[$ the perfect fluid correspondence is always verified. However, the sound speed squared of the UDE fluid
\be
{c}_{\rm s(ude)}^{2}=\frac{p_{\rm ude,\it \widetilde X}}{\rho_{\rm ude, \it\widetilde X}}= w_{\rm ude}\left(1-\frac{m}{\sqrt{2\widetilde{X}}}\right)\,,
\ee
is negative for ${\widetilde X} \in \ ]m^2/2,+\infty[$. Although, this may appear to constitute a no-go condition for this model, that may not be the case. Indeed, for $w_{\rm de}$ sufficiently close to $-1$, the negative sound speed would only become significant in extremely underdense regions (note that $\widetilde{c}_s^2 \to 0$ when $\widetilde{X}^2 \to m^2/2$). In any case, UDE models with a negative sound speed may be avoided by starting with a nonphantom DE model satisfying the condition ${c}_{\rm s(de)}^{2}>0$. 

\subsubsection{Input DE model: Chaplygin gas}

Consider the case of the generalized Chaplygin gas defined by the Lagrangian \cite{2002PhRvD..66d3507B}
\be
\mathcal L (X)=-A^{\frac{1}{1+\alpha}}\left( 1-\left( 2X\right)^{\frac{1+\alpha}{2\alpha}}\right)^{\frac{\alpha}{1+\alpha}}\,,
\ee
where $0 < \alpha < 1$ and $A>0$ are constants (in the following we shall also assume that variables with dimensions of mass are measured in some arbitrary mass unit $m_{\rm unit}$). Although the generalized Chaplygin gas is a UDE prototype, here we shall take it as our input DE model --- the corresponding equation of state parameter and sound speed squared are given, respectively, by
\be
w_{\rm de}=-\frac{A}{\rho^{1+\alpha}} \, \qquad c^2_{\rm s(de)}=-\alpha w_{\rm de}\,,
\ee
with $-1 <w_{\rm de} < 0$ and $0< c_s^2 < 1$ (assuming that $\rho > A^{\frac{1}{1+\alpha}}$). In this case,
\be
\widetilde {\mathcal L} (\widetilde X)=-A^{\frac{1}{1+\alpha}} \xi(\widetilde X)^{\frac{\alpha}{1+\alpha}}\,,
\ee
with
\be
\xi(\widetilde X) = 1-\left(\sqrt {2 \widetilde X} - m\right)^{\frac{1+\alpha}{\alpha}}
\ee
describes a UDE model with proper pressure $p_{\rm ude} (\widetilde X) =  \widetilde {\mathcal L} (\widetilde X)$ and proper energy density
\bq
\rho_{\rm ude} (\widetilde X) &=& 2 \widetilde X \widetilde {\mathcal L}_{,\widetilde X} - \widetilde {\mathcal L} = \rho_{\rm m} (\widetilde X)+\rho_{\rm de} (\widetilde X)\,,\\
\rho_{\rm de} (\widetilde X) &=& A^{\frac{1}{1+\alpha}}\xi(\widetilde X)^{-\frac{1}{1+\alpha}}\,,\\
\rho_{\rm m} (\widetilde X) &=& mn= m\left(\sqrt {2 \widetilde X} - m\right)^{\frac{1}{\alpha}} \rho_{\rm de} \,.
\eq
 At late times $\widetilde X \to m^2/2$, thus implying that both $\rho_{\rm ude}$ and $-p_{\rm ude}$ approach the constant value $A^{1/(1+\alpha)}$. On the other hand, at early times $\widetilde X$ approaches $(m+1/2)^2/2$. As a result, the energy density becomes large and $\rho_{\rm m}$ is roughly proportional to $\rho_{\rm de}$ --- this behavior is explained by the fact that the Chaplygin gas behaves as CDM for densities much greater than $A^{1/(1+\alpha)}$. Notice that for $m$ sufficiently large it is always possible to ensure that $\rho_{\rm ude} \sim \rho_{\rm m}$ at early times. As previously discussed, the positive sound speed squared of the input generalized Chaplygin model implies that ${c}_{\rm s(ude)}^{2}>0$, thus guaranteeing that the resulting UDE model is free from pathological instabilities associated with an imaginary sound speed.

\subsubsection{Restrictions on isentropic UDE models}

Let us now consider the following parametrization of the equation of state of the original DE fluid \cite{2003PhRvL..90i1301L}
\begin{equation}
w_{\rm de}(z)=w_0+\Delta w \frac{z}{1+z}\,,
\end{equation}
where $w_0 \equiv w_{\rm de}(z=0)$, $w_\infty \equiv w_{\rm de}(z=\infty)$ and $\Delta w \equiv w_\infty-w_0$. It is possible to show that this parametrization of $w\left(z\right)$ admits a purely kinetic Lagrangian formulation \cite{2007APh....28..263D}. The energy density of the corresponding UDE fluid is equal to
\begin{equation}
\rho_{\rm ude}=\rho_{\rm ude0}\left[\left(1+z\right)^{3\left(1+w_{\infty}\right)}e^{3\Delta w/(1+z)}+\mathcal{Q}\left(1+z\right)^3\right]\,,
\end{equation}
and the sound speed squared is
\begin{equation}
\label{eq:cs}
c_{\rm s(ude)}^2=\frac{\left(1+w_\infty\right)w_{\rm de}(z) +\left(1-3w_{\rm de}(z)\right) \frac{\Delta w}{ 3(1+z)}}{1+w_{\rm de}(z)+\mathcal{Q}(1+z)^{-3 w_\infty}e^{-3 \Delta w/(1+z)}}\,,
\end{equation}
where $\mathcal{Q} \equiv mn_{0}/\rho_{\rm ude0}$ and $\rho_{\rm ude0} \equiv \rho_{\rm ude} (z=0)$. At the present time
\begin{equation}
c_{\rm s(ude)0}^2=\frac{\Delta w+3w_0\left(1+w_0\right)}{3(1+w_0+\mathcal{Q}e^{-3\Delta w})}\,.
\end{equation}
If one assumes that the original fluid is a DE fluid with $w_0$ sufficiently close to $-1$ one finds
\begin{equation}
c_{\rm s(ude)0}^2=\frac{w_\infty+1}{3\mathcal{Q}}e^{3\left(w_{\infty}+1\right)}\,.
\end{equation}

In order for the transformed fluid to play a UDE role $\mathcal{Q} \sim \Omega_{\rm cdm0}/\Omega_{\rm de0}\sim 3/7$, where $\Omega_{\rm cdm0}$ and $\Omega_{\rm de0}$ are the fractional DM and DE densities inferred from the observations. This in turn implies that $c_{\rm s(ude)0}^2 \sim (w_\infty+1)e^{3\left(w_{\infty}+1\right)}$. Therefore, large sound speeds at recent times would be unavoidable, unless $|w_\infty + 1| \ll 1$. One can estimate how small this value has to be in order to be consistent with the standard growth of perturbation on linear scales by imposing that $c_{\rm s(ude)0} \lesssim 10^{-3}$ \cite{2017PhLB..770..213F}.

Hence the variation of $w$ is limited to $|1+w_{\infty}|\lesssim 10^{-6}$, meaning that the original fluid has to follow very closely the behavior of a cosmological constant. More generally, Eq. \eqref{eq:cs} implies that large sound speeds at low redshifts can be avoided only if both $|w_\infty + 1|$ and $|w_0 + 1|$ are extremely small. Such stringent constraints regarding a non-null sound speed are typical for UDE models as far as linear perturbation theory is concerned \cite{2004PhRvD..69l3524S, 2017PhLB..770..213F} (see also \cite{Avelino:2015dwa}). However, it has been shown that the clustering on nonlinear scales can have a potential impact on the large scale evolution of the Universe, specially in UDE scenarios \cite{2004PhRvD..69d1301A, 2004JCAP...11..008B, 2010CQGra..27q5013R}. Taking into account nonlinear effects may render these models (ruled out in a linear analysis) consistent with cosmological observations \cite{Beca:2005gc, 2013PhRvD..87d3527D, 2014PhRvD..89j3004A}.\\

\section{\label{sec:conc} Conclusions}

In this paper we have investigated the degeneracies between the energy-momentum tensor and the on-shell Lagrangian of a perfect fluid, explicitly showing that one does not univocally determine the other. We have discussed the appropriateness of various Lagrangians to describe the dynamics of different components of the cosmic energy budget, distinguishing those that may be essentially modeled as a collection of point particles, such as baryons, photons or neutrinos, from those that do not, such as DE. We have explicitly shown that aforementioned distinction is particularly relevant if a NMC exists with the gravitational field or other matter fields, in which case the knowledge of the on-shell Lagrangian can be essential to compute the overall dynamics. This point has been overlooked in the literature, where it is often wrongly assumed that there is a freedom of choice of the on-shell Lagrangian, even when describing standard model particles.

We have also explored the fact that models with the same on-shell Lagrangian may have different proper energy densities. We have used this result to establish a map between DE models described by purely kinetic Lagrangians and UDE models, characterizing the correspondence between their equation of state and sound speed parameters. Successful UDE models are essentially required to match the observed evolution of the proper pressure at low redshifts, while, at the same time, accounting for the observed large scale structure of the Universe. The simplest way to accomplish this, followed in Secs. \ref{sec:Model}B and \ref{sec:Model}C, is to combine DM and DE into a single perfect fluid --- i.e., a perfect fluid with proper pressure equal to the observed proper pressure (usually attributed to the DE) and proper density approximately equal to $\sim 95 \%$ of the energy density of the Universe at the present time (thus accounting for both the CDM and DE energy densities). This allows one to map DE into UDE and to build well-defined models beyond $\Lambda \rm CDM$ which can be confronted with observations. Furthermore, we have shown that the sound speed squared of the resulting UDE models are always positive, as long as that is also verified in the case of the input nonphantom DE models, thus ensuring the avoidance of pathological instabilities at a linear level --- notice that at the nonlinear level this is guaranteed by the fact that if the density is large, UDE behaves essentially as CDM. We have also briefly discussed the linear sound speed problem of UDE models as well as a possible way out associated with their nonlinear dynamics, arguing that, depending on the level of nonlinear clustering, they may turn out to be compatible with observations.

\begin{acknowledgments}

V.M.C.F. was supported by the FCT fellowship (PD/BD/135229/2017), within the FCT PhD Program PhD::SPACE (PD/00040/2012). P.P.A. acknowledges the support from Fundação para a Ciência e a Tecnologia (FCT) through the Sabbatical Grant No. SFRH/BSAB/150322/2019. R.P.L.A. was supported by the FCT fellowship SFRH/BD/132546/2017. This work was also supported by FCT through national funds (PTDC/FIS-PAR/31938/2017) and by FEDER—Fundo Europeu de Desenvolvimento Regional through COMPETE2020 - Programa Operacional Competitividade e Internacionaliza{\c c}\~ao (POCI-01-0145-FEDER-031938). Funding of this work has also been provided through the research Grants No. UID/FIS/04434/2019, No. UIDB/04434/2020 and UIDP/04434/2020.

\end{acknowledgments}


\bibliography{Lagrangian_perfect_fluid}

\begin{thebibliography}{110}
\expandafter\ifx\csname natexlab\endcsname\relax\def\natexlab#1{#1}\fi
\expandafter\ifx\csname bibnamefont\endcsname\relax
  \def\bibnamefont#1{#1}\fi
\expandafter\ifx\csname bibfnamefont\endcsname\relax
  \def\bibfnamefont#1{#1}\fi
\expandafter\ifx\csname citenamefont\endcsname\relax
  \def\citenamefont#1{#1}\fi
\expandafter\ifx\csname url\endcsname\relax
  \def\url#1{\texttt{#1}}\fi
\expandafter\ifx\csname urlprefix\endcsname\relax\def\urlprefix{URL }\fi
\providecommand{\bibinfo}[2]{#2}
\providecommand{\eprint}[2][]{\url{#2}}

\bibitem[{\citenamefont{Aad et~al.}(2012)}]{Aad:2012tfa}
\bibinfo{author}{\bibfnamefont{G.}~\bibnamefont{Aad}} \bibnamefont{et~al.}
  (\bibinfo{collaboration}{ATLAS}), \bibinfo{journal}{Phys. Lett.}
  \textbf{\bibinfo{volume}{B716}}, \bibinfo{pages}{1} (\bibinfo{year}{2012}),
  \eprint{1207.7214}.

\bibitem[{\citenamefont{Chatrchyan et~al.}(2012)}]{Chatrchyan:2012xdj}
\bibinfo{author}{\bibfnamefont{S.}~\bibnamefont{Chatrchyan}}
  \bibnamefont{et~al.} (\bibinfo{collaboration}{CMS}), \bibinfo{journal}{Phys.
  Lett.} \textbf{\bibinfo{volume}{B716}}, \bibinfo{pages}{30}
  (\bibinfo{year}{2012}), \eprint{1207.7235}.

\bibitem[{\citenamefont{{Linde}}(1983)}]{1983PhLB..129..177L}
\bibinfo{author}{\bibfnamefont{A.~D.} \bibnamefont{{Linde}}},
  \bibinfo{journal}{Physics Letters B} \textbf{\bibinfo{volume}{129}},
  \bibinfo{pages}{177} (\bibinfo{year}{1983}).

\bibitem[{\citenamefont{{Lyth} and {Riotto}}(1999)}]{1999PhR...314....1L}
\bibinfo{author}{\bibfnamefont{D.~H. D.~H.} \bibnamefont{{Lyth}}}
  \bibnamefont{and} \bibinfo{author}{\bibfnamefont{A.~A.}
  \bibnamefont{{Riotto}}}, \bibinfo{journal}{Phys. Rept.}
  \textbf{\bibinfo{volume}{314}}, \bibinfo{pages}{1} (\bibinfo{year}{1999}),
  \eprint{hep-ph/9807278}.

\bibitem[{\citenamefont{{Armend{\'a}riz-Pic{\'o}n}
  et~al.}(1999)\citenamefont{{Armend{\'a}riz-Pic{\'o}n}, {Damour}, and
  {Mukhanov}}}]{1999PhLB..458..209A}
\bibinfo{author}{\bibfnamefont{C.}~\bibnamefont{{Armend{\'a}riz-Pic{\'o}n}}},
  \bibinfo{author}{\bibfnamefont{T.}~\bibnamefont{{Damour}}}, \bibnamefont{and}
  \bibinfo{author}{\bibfnamefont{V.}~\bibnamefont{{Mukhanov}}},
  \bibinfo{journal}{Physics Letters B} \textbf{\bibinfo{volume}{458}},
  \bibinfo{pages}{209} (\bibinfo{year}{1999}), \eprint{hep-th/9904075}.

\bibitem[{\citenamefont{{D{\'\i}ez-Tejedor} and
  {Feinstein}}(2006)}]{2006PhLA..350..315D}
\bibinfo{author}{\bibfnamefont{A.}~\bibnamefont{{D{\'\i}ez-Tejedor}}}
  \bibnamefont{and}
  \bibinfo{author}{\bibfnamefont{A.}~\bibnamefont{{Feinstein}}},
  \bibinfo{journal}{Physics Letters A} \textbf{\bibinfo{volume}{350}},
  \bibinfo{pages}{315} (\bibinfo{year}{2006}), \eprint{gr-qc/0505105}.

\bibitem[{\citenamefont{{Allahverdi} et~al.}(2010)\citenamefont{{Allahverdi},
  {Brandenberger}, {Cyr-Racine}, and {Mazumdar}}}]{2010ARNPS..60...27A}
\bibinfo{author}{\bibfnamefont{R.}~\bibnamefont{{Allahverdi}}},
  \bibinfo{author}{\bibfnamefont{R.}~\bibnamefont{{Brandenberger}}},
  \bibinfo{author}{\bibfnamefont{F.-Y.} \bibnamefont{{Cyr-Racine}}},
  \bibnamefont{and}
  \bibinfo{author}{\bibfnamefont{A.}~\bibnamefont{{Mazumdar}}},
  \bibinfo{journal}{Annual Review of Nuclear and Particle Science}
  \textbf{\bibinfo{volume}{60}}, \bibinfo{pages}{27} (\bibinfo{year}{2010}),
  \eprint{1001.2600}.

\bibitem[{\citenamefont{{Riess} et~al.}(1998)\citenamefont{{Riess},
  {Filippenko}, {Challis}, {Clocchiatti}, {Diercks}, {Garnavich}, {Gilliland},
  {Hogan}, {Jha}, {Kirshner} et~al.}}]{1998AJ....116.1009R}
\bibinfo{author}{\bibfnamefont{A.~G.} \bibnamefont{{Riess}}},
  \bibinfo{author}{\bibfnamefont{A.~V.} \bibnamefont{{Filippenko}}},
  \bibinfo{author}{\bibfnamefont{P.}~\bibnamefont{{Challis}}},
  \bibinfo{author}{\bibfnamefont{A.}~\bibnamefont{{Clocchiatti}}},
  \bibinfo{author}{\bibfnamefont{A.}~\bibnamefont{{Diercks}}},
  \bibinfo{author}{\bibfnamefont{P.~M.} \bibnamefont{{Garnavich}}},
  \bibinfo{author}{\bibfnamefont{R.~L.} \bibnamefont{{Gilliland}}},
  \bibinfo{author}{\bibfnamefont{C.~J.} \bibnamefont{{Hogan}}},
  \bibinfo{author}{\bibfnamefont{S.}~\bibnamefont{{Jha}}},
  \bibinfo{author}{\bibfnamefont{R.~P.} \bibnamefont{{Kirshner}}},
  \bibnamefont{et~al.}, \bibinfo{journal}{Astronomical Journal}
  \textbf{\bibinfo{volume}{116}}, \bibinfo{pages}{1009} (\bibinfo{year}{1998}),
  \eprint{astro-ph/9805201}.

\bibitem[{\citenamefont{{Perlmutter} et~al.}(1999)\citenamefont{{Perlmutter},
  {Aldering}, {Goldhaber}, {Knop}, {Nugent}, {Castro}, {Deustua}, {Fabbro},
  {Goobar}, {Groom} et~al.}}]{1999ApJ...517..565P}
\bibinfo{author}{\bibfnamefont{S.}~\bibnamefont{{Perlmutter}}},
  \bibinfo{author}{\bibfnamefont{G.}~\bibnamefont{{Aldering}}},
  \bibinfo{author}{\bibfnamefont{G.}~\bibnamefont{{Goldhaber}}},
  \bibinfo{author}{\bibfnamefont{R.~A.} \bibnamefont{{Knop}}},
  \bibinfo{author}{\bibfnamefont{P.}~\bibnamefont{{Nugent}}},
  \bibinfo{author}{\bibfnamefont{P.~G.} \bibnamefont{{Castro}}},
  \bibinfo{author}{\bibfnamefont{S.}~\bibnamefont{{Deustua}}},
  \bibinfo{author}{\bibfnamefont{S.}~\bibnamefont{{Fabbro}}},
  \bibinfo{author}{\bibfnamefont{A.}~\bibnamefont{{Goobar}}},
  \bibinfo{author}{\bibfnamefont{D.~E.} \bibnamefont{{Groom}}},
  \bibnamefont{et~al.}, \bibinfo{journal}{\apj} \textbf{\bibinfo{volume}{517}},
  \bibinfo{pages}{565} (\bibinfo{year}{1999}), \eprint{astro-ph/9812133}.

\bibitem[{\citenamefont{{Armendariz-Picon}
  et~al.}(2001)\citenamefont{{Armendariz-Picon}, {Mukhanov}, and
  {Steinhardt}}}]{2001PhRvD..63j3510A}
\bibinfo{author}{\bibfnamefont{C.}~\bibnamefont{{Armendariz-Picon}}},
  \bibinfo{author}{\bibfnamefont{V.}~\bibnamefont{{Mukhanov}}},
  \bibnamefont{and} \bibinfo{author}{\bibfnamefont{P.~J.}
  \bibnamefont{{Steinhardt}}}, \bibinfo{journal}{\prd}
  \textbf{\bibinfo{volume}{63}}, \bibinfo{eid}{103510} (\bibinfo{year}{2001}),
  \eprint{astro-ph/0006373}.

\bibitem[{\citenamefont{{Copeland} et~al.}(2006)\citenamefont{{Copeland},
  {Sami}, and {Tsujikawa}}}]{2006IJMPD..15.1753C}
\bibinfo{author}{\bibfnamefont{E.~J.} \bibnamefont{{Copeland}}},
  \bibinfo{author}{\bibfnamefont{M.}~\bibnamefont{{Sami}}}, \bibnamefont{and}
  \bibinfo{author}{\bibfnamefont{S.}~\bibnamefont{{Tsujikawa}}},
  \bibinfo{journal}{International Journal of Modern Physics D}
  \textbf{\bibinfo{volume}{15}}, \bibinfo{pages}{1753} (\bibinfo{year}{2006}),
  \eprint{hep-th/0603057}.

\bibitem[{\citenamefont{{Kang} et~al.}(2007)\citenamefont{{Kang}, {Vanchurin},
  and {Winitzki}}}]{2007PhRvD..76h3511K}
\bibinfo{author}{\bibfnamefont{J.~U.} \bibnamefont{{Kang}}},
  \bibinfo{author}{\bibfnamefont{V.}~\bibnamefont{{Vanchurin}}},
  \bibnamefont{and}
  \bibinfo{author}{\bibfnamefont{S.}~\bibnamefont{{Winitzki}}},
  \bibinfo{journal}{\prd} \textbf{\bibinfo{volume}{76}}, \bibinfo{eid}{083511}
  (\bibinfo{year}{2007}), \eprint{0706.3994}.

\bibitem[{\citenamefont{{Gao} et~al.}(2011)\citenamefont{{Gao}, {Gong}, {Wang},
  and {Chen}}}]{2011PhLB..702..107G}
\bibinfo{author}{\bibfnamefont{C.}~\bibnamefont{{Gao}}},
  \bibinfo{author}{\bibfnamefont{Y.}~\bibnamefont{{Gong}}},
  \bibinfo{author}{\bibfnamefont{X.}~\bibnamefont{{Wang}}}, \bibnamefont{and}
  \bibinfo{author}{\bibfnamefont{X.}~\bibnamefont{{Chen}}},
  \bibinfo{journal}{Physics Letters B} \textbf{\bibinfo{volume}{702}},
  \bibinfo{pages}{107} (\bibinfo{year}{2011}), \eprint{1003.6056}.

\bibitem[{\citenamefont{{Hui} et~al.}(2017)\citenamefont{{Hui}, {Ostriker},
  {Tremaine}, and {Witten}}}]{2017PhRvD..95d3541H}
\bibinfo{author}{\bibfnamefont{L.}~\bibnamefont{{Hui}}},
  \bibinfo{author}{\bibfnamefont{J.~P.} \bibnamefont{{Ostriker}}},
  \bibinfo{author}{\bibfnamefont{S.}~\bibnamefont{{Tremaine}}},
  \bibnamefont{and} \bibinfo{author}{\bibfnamefont{E.}~\bibnamefont{{Witten}}},
  \bibinfo{journal}{\prd} \textbf{\bibinfo{volume}{95}}, \bibinfo{eid}{043541}
  (\bibinfo{year}{2017}), \eprint{1610.08297}.

\bibitem[{\citenamefont{Ure\~{n}a L\'{o}pez}(2019)}]{Urena-Lopez:2019kud}
\bibinfo{author}{\bibfnamefont{L.~A.} \bibnamefont{Ure\~{n}a L\'{o}pez}},
  \bibinfo{journal}{Front. Astron. Space Sci.} \textbf{\bibinfo{volume}{6}},
  \bibinfo{pages}{47} (\bibinfo{year}{2019}).

\bibitem[{\citenamefont{{Clifton} et~al.}(2012)\citenamefont{{Clifton},
  {Ferreira}, {Padilla}, and {Skordis}}}]{2012PhR...513....1C}
\bibinfo{author}{\bibfnamefont{T.}~\bibnamefont{{Clifton}}},
  \bibinfo{author}{\bibfnamefont{P.~G.} \bibnamefont{{Ferreira}}},
  \bibinfo{author}{\bibfnamefont{A.}~\bibnamefont{{Padilla}}},
  \bibnamefont{and}
  \bibinfo{author}{\bibfnamefont{C.}~\bibnamefont{{Skordis}}},
  \bibinfo{journal}{Phys. Rept.} \textbf{\bibinfo{volume}{513}},
  \bibinfo{pages}{1} (\bibinfo{year}{2012}), \eprint{1106.2476}.

\bibitem[{\citenamefont{Bamba et~al.}(2012{\natexlab{a}})\citenamefont{Bamba,
  Capozziello, Nojiri, and Odintsov}}]{Bamba:2012cp}
\bibinfo{author}{\bibfnamefont{K.}~\bibnamefont{Bamba}},
  \bibinfo{author}{\bibfnamefont{S.}~\bibnamefont{Capozziello}},
  \bibinfo{author}{\bibfnamefont{S.}~\bibnamefont{Nojiri}}, \bibnamefont{and}
  \bibinfo{author}{\bibfnamefont{S.~D.} \bibnamefont{Odintsov}},
  \bibinfo{journal}{Astrophys. Space Sci.} \textbf{\bibinfo{volume}{342}},
  \bibinfo{pages}{155} (\bibinfo{year}{2012}{\natexlab{a}}),
  \eprint{1205.3421}.

\bibitem[{\citenamefont{{Joyce} et~al.}(2015)\citenamefont{{Joyce}, {Jain},
  {Khoury}, and {Trodden}}}]{2015PhR...568....1J}
\bibinfo{author}{\bibfnamefont{A.}~\bibnamefont{{Joyce}}},
  \bibinfo{author}{\bibfnamefont{B.}~\bibnamefont{{Jain}}},
  \bibinfo{author}{\bibfnamefont{J.}~\bibnamefont{{Khoury}}}, \bibnamefont{and}
  \bibinfo{author}{\bibfnamefont{M.}~\bibnamefont{{Trodden}}},
  \bibinfo{journal}{Phys. Rept.} \textbf{\bibinfo{volume}{568}},
  \bibinfo{pages}{1} (\bibinfo{year}{2015}), \eprint{1407.0059}.

\bibitem[{\citenamefont{{Avelino} et~al.}(2016)\citenamefont{{Avelino},
  {Barreiro}, {Carvalho}, {da Silva}, {Lobo}, {Martin-Moruno}, {Mimoso},
  {Nunes}, {Rubiera-Garcia}, {Saez-Gomez} et~al.}}]{2016arXiv160702979A}
\bibinfo{author}{\bibfnamefont{P.~P.} \bibnamefont{{Avelino}}},
  \bibinfo{author}{\bibfnamefont{T.}~\bibnamefont{{Barreiro}}},
  \bibinfo{author}{\bibfnamefont{C.~S.} \bibnamefont{{Carvalho}}},
  \bibinfo{author}{\bibfnamefont{A.}~\bibnamefont{{da Silva}}},
  \bibinfo{author}{\bibfnamefont{F.~S.~N.} \bibnamefont{{Lobo}}},
  \bibinfo{author}{\bibfnamefont{P.}~\bibnamefont{{Martin-Moruno}}},
  \bibinfo{author}{\bibfnamefont{J.~P.} \bibnamefont{{Mimoso}}},
  \bibinfo{author}{\bibfnamefont{N.~J.} \bibnamefont{{Nunes}}},
  \bibinfo{author}{\bibfnamefont{D.}~\bibnamefont{{Rubiera-Garcia}}},
  \bibinfo{author}{\bibfnamefont{D.}~\bibnamefont{{Saez-Gomez}}},
  \bibnamefont{et~al.}, \bibinfo{journal}{arXiv e-prints}
  \bibinfo{eid}{arXiv:1607.02979} (\bibinfo{year}{2016}), \eprint{1607.02979}.

\bibitem[{\citenamefont{{Hossain} et~al.}(2015)\citenamefont{{Hossain},
  {Myrzakulov}, {Sami}, and {Saridakis}}}]{2015IJMPD..2430014H}
\bibinfo{author}{\bibfnamefont{M.~W.} \bibnamefont{{Hossain}}},
  \bibinfo{author}{\bibfnamefont{R.}~\bibnamefont{{Myrzakulov}}},
  \bibinfo{author}{\bibfnamefont{M.}~\bibnamefont{{Sami}}}, \bibnamefont{and}
  \bibinfo{author}{\bibfnamefont{E.~N.} \bibnamefont{{Saridakis}}},
  \bibinfo{journal}{International Journal of Modern Physics D}
  \textbf{\bibinfo{volume}{24}}, \bibinfo{eid}{1530014} (\bibinfo{year}{2015}),
  \eprint{1410.6100}.

\bibitem[{\citenamefont{{Kamenshchik} et~al.}(2001)\citenamefont{{Kamenshchik},
  {Moschella}, and {Pasquier}}}]{2001PhLB..511..265K}
\bibinfo{author}{\bibfnamefont{A.}~\bibnamefont{{Kamenshchik}}},
  \bibinfo{author}{\bibfnamefont{U.}~\bibnamefont{{Moschella}}},
  \bibnamefont{and}
  \bibinfo{author}{\bibfnamefont{V.}~\bibnamefont{{Pasquier}}},
  \bibinfo{journal}{Physics Letters B} \textbf{\bibinfo{volume}{511}},
  \bibinfo{pages}{265} (\bibinfo{year}{2001}), \eprint{gr-qc/0103004}.

\bibitem[{\citenamefont{{Bili{\'c}} et~al.}(2002)\citenamefont{{Bili{\'c}},
  {Tupper}, and {Viollier}}}]{2002PhLB..535...17B}
\bibinfo{author}{\bibfnamefont{N.}~\bibnamefont{{Bili{\'c}}}},
  \bibinfo{author}{\bibfnamefont{G.~B.} \bibnamefont{{Tupper}}},
  \bibnamefont{and} \bibinfo{author}{\bibfnamefont{R.~D.}
  \bibnamefont{{Viollier}}}, \bibinfo{journal}{Physics Letters B}
  \textbf{\bibinfo{volume}{535}}, \bibinfo{pages}{17} (\bibinfo{year}{2002}),
  \eprint{astro-ph/0111325}.

\bibitem[{\citenamefont{{Padmanabhan} and
  {Choudhury}}(2002)}]{2002PhRvD..66h1301P}
\bibinfo{author}{\bibfnamefont{T.}~\bibnamefont{{Padmanabhan}}}
  \bibnamefont{and} \bibinfo{author}{\bibfnamefont{T.~R.}
  \bibnamefont{{Choudhury}}}, \bibinfo{journal}{\prd}
  \textbf{\bibinfo{volume}{66}}, \bibinfo{eid}{081301} (\bibinfo{year}{2002}),
  \eprint{hep-th/0205055}.

\bibitem[{\citenamefont{{Bento} et~al.}(2002)\citenamefont{{Bento},
  {Bertolami}, and {Sen}}}]{2002PhRvD..66d3507B}
\bibinfo{author}{\bibfnamefont{M.~C.} \bibnamefont{{Bento}}},
  \bibinfo{author}{\bibfnamefont{O.}~\bibnamefont{{Bertolami}}},
  \bibnamefont{and} \bibinfo{author}{\bibfnamefont{A.~A.} \bibnamefont{{Sen}}},
  \bibinfo{journal}{\prd} \textbf{\bibinfo{volume}{66}}, \bibinfo{eid}{043507}
  (\bibinfo{year}{2002}), \eprint{gr-qc/0202064}.

\bibitem[{\citenamefont{{Scherrer}}(2004)}]{2004PhRvL..93a1301S}
\bibinfo{author}{\bibfnamefont{R.~J.} \bibnamefont{{Scherrer}}},
  \bibinfo{journal}{Physical Review Letters} \textbf{\bibinfo{volume}{93}},
  \bibinfo{eid}{011301} (\bibinfo{year}{2004}), \eprint{astro-ph/0402316}.

\bibitem[{\citenamefont{{Chimento} et~al.}(2005)\citenamefont{{Chimento},
  {Forte}, and {Lazkoz}}}]{2005MPLA...20.2075C}
\bibinfo{author}{\bibfnamefont{L.~P.} \bibnamefont{{Chimento}}},
  \bibinfo{author}{\bibfnamefont{M.}~\bibnamefont{{Forte}}}, \bibnamefont{and}
  \bibinfo{author}{\bibfnamefont{R.}~\bibnamefont{{Lazkoz}}},
  \bibinfo{journal}{Modern Physics Letters A} \textbf{\bibinfo{volume}{20}},
  \bibinfo{pages}{2075} (\bibinfo{year}{2005}), \eprint{astro-ph/0407288}.

\bibitem[{\citenamefont{{Bertacca} et~al.}(2007)\citenamefont{{Bertacca},
  {Matarrese}, and {Pietroni}}}]{2007MPLA...22.2893B}
\bibinfo{author}{\bibfnamefont{D.}~\bibnamefont{{Bertacca}}},
  \bibinfo{author}{\bibfnamefont{S.}~\bibnamefont{{Matarrese}}},
  \bibnamefont{and}
  \bibinfo{author}{\bibfnamefont{M.}~\bibnamefont{{Pietroni}}},
  \bibinfo{journal}{Modern Physics Letters A} \textbf{\bibinfo{volume}{22}},
  \bibinfo{pages}{2893} (\bibinfo{year}{2007}), \eprint{astro-ph/0703259}.

\bibitem[{\citenamefont{{Lim} et~al.}(2010)\citenamefont{{Lim}, {Sawicki}, and
  {Vikman}}}]{2010JCAP...05..012L}
\bibinfo{author}{\bibfnamefont{E.~A.} \bibnamefont{{Lim}}},
  \bibinfo{author}{\bibfnamefont{I.}~\bibnamefont{{Sawicki}}},
  \bibnamefont{and} \bibinfo{author}{\bibfnamefont{A.}~\bibnamefont{{Vikman}}},
  \bibinfo{journal}{Journal of Cosmology and Astroparticle Physics}
  \textbf{\bibinfo{volume}{2010}}, \bibinfo{eid}{012} (\bibinfo{year}{2010}),
  \eprint{1003.5751}.

\bibitem[{\citenamefont{{Ferreira} and {Avelino}}(2018)}]{2018PhRvD..98d3515F}
\bibinfo{author}{\bibfnamefont{V.~M.~C.} \bibnamefont{{Ferreira}}}
  \bibnamefont{and} \bibinfo{author}{\bibfnamefont{P.~P.}
  \bibnamefont{{Avelino}}}, \bibinfo{journal}{\prd}
  \textbf{\bibinfo{volume}{98}}, \bibinfo{eid}{043515} (\bibinfo{year}{2018}),
  \eprint{1807.04656}.

\bibitem[{\citenamefont{{Luongo} and {Muccino}}(2018)}]{2018PhRvD..98j3520L}
\bibinfo{author}{\bibfnamefont{O.}~\bibnamefont{{Luongo}}} \bibnamefont{and}
  \bibinfo{author}{\bibfnamefont{M.}~\bibnamefont{{Muccino}}},
  \bibinfo{journal}{\prd} \textbf{\bibinfo{volume}{98}}, \bibinfo{eid}{103520}
  (\bibinfo{year}{2018}), \eprint{1807.00180}.

\bibitem[{\citenamefont{Boshkayev et~al.}(2019)\citenamefont{Boshkayev,
  D'Agostino, and Luongo}}]{Boshkayev:2019qcx}
\bibinfo{author}{\bibfnamefont{K.}~\bibnamefont{Boshkayev}},
  \bibinfo{author}{\bibfnamefont{R.}~\bibnamefont{D'Agostino}},
  \bibnamefont{and} \bibinfo{author}{\bibfnamefont{O.}~\bibnamefont{Luongo}},
  \bibinfo{journal}{Eur. Phys. J. C} \textbf{\bibinfo{volume}{79}},
  \bibinfo{pages}{332} (\bibinfo{year}{2019}), \eprint{1901.01031}.

\bibitem[{\citenamefont{{Liddle} and
  {Ure{\~n}a-L{\'o}pez}}(2006)}]{2006PhRvL..97p1301L}
\bibinfo{author}{\bibfnamefont{A.~R.} \bibnamefont{{Liddle}}} \bibnamefont{and}
  \bibinfo{author}{\bibfnamefont{L.~A.} \bibnamefont{{Ure{\~n}a-L{\'o}pez}}},
  \bibinfo{journal}{\prl} \textbf{\bibinfo{volume}{97}}, \bibinfo{eid}{161301}
  (\bibinfo{year}{2006}), \eprint{astro-ph/0605205}.

\bibitem[{\citenamefont{{Bose} and {Majumdar}}(2009)}]{2009PhRvD..79j3517B}
\bibinfo{author}{\bibfnamefont{N.}~\bibnamefont{{Bose}}} \bibnamefont{and}
  \bibinfo{author}{\bibfnamefont{A.~S.} \bibnamefont{{Majumdar}}},
  \bibinfo{journal}{\prd} \textbf{\bibinfo{volume}{79}}, \bibinfo{eid}{103517}
  (\bibinfo{year}{2009}), \eprint{0812.4131}.

\bibitem[{\citenamefont{{de-Santiago} and
  {Cervantes-Cota}}(2011)}]{2011PhRvD..83f3502D}
\bibinfo{author}{\bibfnamefont{J.}~\bibnamefont{{de-Santiago}}}
  \bibnamefont{and} \bibinfo{author}{\bibfnamefont{J.~L.}
  \bibnamefont{{Cervantes-Cota}}}, \bibinfo{journal}{\prd}
  \textbf{\bibinfo{volume}{83}}, \bibinfo{eid}{063502} (\bibinfo{year}{2011}),
  \eprint{1102.1777}.

\bibitem[{\citenamefont{Madsen}(1988)}]{Madsen:1988ph}
\bibinfo{author}{\bibfnamefont{M.~S.} \bibnamefont{Madsen}},
  \bibinfo{journal}{Class. Quant. Grav.} \textbf{\bibinfo{volume}{5}},
  \bibinfo{pages}{627} (\bibinfo{year}{1988}).

\bibitem[{\citenamefont{{Arroja} and {Sasaki}}(2010)}]{2010PhRvD..81j7301A}
\bibinfo{author}{\bibfnamefont{F.}~\bibnamefont{{Arroja}}} \bibnamefont{and}
  \bibinfo{author}{\bibfnamefont{M.}~\bibnamefont{{Sasaki}}},
  \bibinfo{journal}{\prd} \textbf{\bibinfo{volume}{81}}, \bibinfo{eid}{107301}
  (\bibinfo{year}{2010}), \eprint{1002.1376}.

\bibitem[{\citenamefont{{Unnikrishnan} and
  {Sriramkumar}}(2010)}]{2010PhRvD..81j3511U}
\bibinfo{author}{\bibfnamefont{S.}~\bibnamefont{{Unnikrishnan}}}
  \bibnamefont{and}
  \bibinfo{author}{\bibfnamefont{L.}~\bibnamefont{{Sriramkumar}}},
  \bibinfo{journal}{\prd} \textbf{\bibinfo{volume}{81}}, \bibinfo{eid}{103511}
  (\bibinfo{year}{2010}), \eprint{1002.0820}.

\bibitem[{\citenamefont{{Piattella} et~al.}(2014)\citenamefont{{Piattella},
  {Fabris}, and {Bili{\'c}}}}]{2014CQGra..31e5006P}
\bibinfo{author}{\bibfnamefont{O.~F.} \bibnamefont{{Piattella}}},
  \bibinfo{author}{\bibfnamefont{J.~C.} \bibnamefont{{Fabris}}},
  \bibnamefont{and}
  \bibinfo{author}{\bibfnamefont{N.}~\bibnamefont{{Bili{\'c}}}},
  \bibinfo{journal}{Classical and Quantum Gravity}
  \textbf{\bibinfo{volume}{31}}, \bibinfo{eid}{055006} (\bibinfo{year}{2014}),
  \eprint{1309.4282}.

\bibitem[{\citenamefont{Schutz}(1970)}]{PhysRevD.2.2762}
\bibinfo{author}{\bibfnamefont{B.~F.} \bibnamefont{Schutz}},
  \bibinfo{journal}{Phys. Rev. D} \textbf{\bibinfo{volume}{2}},
  \bibinfo{pages}{2762} (\bibinfo{year}{1970}).

\bibitem[{\citenamefont{Ray}(1972)}]{doi:10.1063/1.1665861}
\bibinfo{author}{\bibfnamefont{J.~R.} \bibnamefont{Ray}},
  \bibinfo{journal}{Journal of Mathematical Physics}
  \textbf{\bibinfo{volume}{13}}, \bibinfo{pages}{1451} (\bibinfo{year}{1972}),
  \eprint{https://doi.org/10.1063/1.1665861}.

\bibitem[{\citenamefont{{Schutz} and {Sorkin}}(1977)}]{1977AnPhy.107....1S}
\bibinfo{author}{\bibfnamefont{B.~F.} \bibnamefont{{Schutz}}} \bibnamefont{and}
  \bibinfo{author}{\bibfnamefont{R.}~\bibnamefont{{Sorkin}}},
  \bibinfo{journal}{Annals of Physics} \textbf{\bibinfo{volume}{107}},
  \bibinfo{pages}{1} (\bibinfo{year}{1977}).

\bibitem[{\citenamefont{{Taub}}(1978)}]{1978AnRFM..10..301T}
\bibinfo{author}{\bibfnamefont{A.~H.} \bibnamefont{{Taub}}},
  \bibinfo{journal}{Annual Review of Fluid Mechanics}
  \textbf{\bibinfo{volume}{10}}, \bibinfo{pages}{301} (\bibinfo{year}{1978}).

\bibitem[{\citenamefont{Matarrese}(1985)}]{Matarrese:1984zw}
\bibinfo{author}{\bibfnamefont{S.}~\bibnamefont{Matarrese}},
  \bibinfo{journal}{Proc. Roy. Soc. Lond.} \textbf{\bibinfo{volume}{A401}},
  \bibinfo{pages}{53} (\bibinfo{year}{1985}).

\bibitem[{\citenamefont{Brown}(1993)}]{Brown:1992kc}
\bibinfo{author}{\bibfnamefont{J.~D.} \bibnamefont{Brown}},
  \bibinfo{journal}{Class. Quant. Grav.} \textbf{\bibinfo{volume}{10}},
  \bibinfo{pages}{1579} (\bibinfo{year}{1993}), \eprint{gr-qc/9304026}.

\bibitem[{\citenamefont{Andersson and Comer}(2007)}]{Andersson:2006nr}
\bibinfo{author}{\bibfnamefont{N.}~\bibnamefont{Andersson}} \bibnamefont{and}
  \bibinfo{author}{\bibfnamefont{G.}~\bibnamefont{Comer}},
  \bibinfo{journal}{Living Rev. Rel.} \textbf{\bibinfo{volume}{10}},
  \bibinfo{pages}{1} (\bibinfo{year}{2007}), \eprint{gr-qc/0605010}.

\bibitem[{\citenamefont{{Minazzoli} and {Harko}}(2012)}]{2012PhRvD..86h7502M}
\bibinfo{author}{\bibfnamefont{O.}~\bibnamefont{{Minazzoli}}} \bibnamefont{and}
  \bibinfo{author}{\bibfnamefont{T.}~\bibnamefont{{Harko}}},
  \bibinfo{journal}{\prd} \textbf{\bibinfo{volume}{86}}, \bibinfo{eid}{087502}
  (\bibinfo{year}{2012}), \eprint{1209.2754}.

\bibitem[{\citenamefont{Avelino and Azevedo}(2018)}]{Avelino2018}
\bibinfo{author}{\bibfnamefont{P.~P.} \bibnamefont{Avelino}} \bibnamefont{and}
  \bibinfo{author}{\bibfnamefont{R.~P.~L.} \bibnamefont{Azevedo}},
  \bibinfo{journal}{Phys. Rev. D} \textbf{\bibinfo{volume}{97}},
  \bibinfo{pages}{64018} (\bibinfo{year}{2018}).

\bibitem[{\citenamefont{Avelino and Sousa}(2018)}]{Avelino2018a}
\bibinfo{author}{\bibfnamefont{P.~P.} \bibnamefont{Avelino}} \bibnamefont{and}
  \bibinfo{author}{\bibfnamefont{L.}~\bibnamefont{Sousa}},
  \bibinfo{journal}{Phys. Rev. D} \textbf{\bibinfo{volume}{97}},
  \bibinfo{pages}{64019} (\bibinfo{year}{2018}), \eprint{1802.03961}.

\bibitem[{\citenamefont{Bertolami et~al.}(2007)\citenamefont{Bertolami,
  Boehmer, Harko, and Lobo}}]{Bertolami:2007gv}
\bibinfo{author}{\bibfnamefont{O.}~\bibnamefont{Bertolami}},
  \bibinfo{author}{\bibfnamefont{C.~G.} \bibnamefont{Boehmer}},
  \bibinfo{author}{\bibfnamefont{T.}~\bibnamefont{Harko}}, \bibnamefont{and}
  \bibinfo{author}{\bibfnamefont{F.~S.} \bibnamefont{Lobo}},
  \bibinfo{journal}{Phys. Rev. D} \textbf{\bibinfo{volume}{75}},
  \bibinfo{pages}{104016} (\bibinfo{year}{2007}), \eprint{0704.1733}.

\bibitem[{\citenamefont{Bertolami et~al.}(2008)\citenamefont{Bertolami, Lobo,
  and Paramos}}]{Bertolami:2008ab}
\bibinfo{author}{\bibfnamefont{O.}~\bibnamefont{Bertolami}},
  \bibinfo{author}{\bibfnamefont{F.~S.} \bibnamefont{Lobo}}, \bibnamefont{and}
  \bibinfo{author}{\bibfnamefont{J.}~\bibnamefont{Paramos}},
  \bibinfo{journal}{Phys. Rev. D} \textbf{\bibinfo{volume}{78}},
  \bibinfo{pages}{064036} (\bibinfo{year}{2008}), \eprint{0806.4434}.

\bibitem[{\citenamefont{Sotiriou and Faraoni}(2008)}]{Sotiriou:2008it}
\bibinfo{author}{\bibfnamefont{T.~P.} \bibnamefont{Sotiriou}} \bibnamefont{and}
  \bibinfo{author}{\bibfnamefont{V.}~\bibnamefont{Faraoni}},
  \bibinfo{journal}{Class. Quant. Grav.} \textbf{\bibinfo{volume}{25}},
  \bibinfo{pages}{205002} (\bibinfo{year}{2008}), \eprint{0805.1249}.

\bibitem[{\citenamefont{Faraoni}(2009)}]{Faraoni:2009rk}
\bibinfo{author}{\bibfnamefont{V.}~\bibnamefont{Faraoni}},
  \bibinfo{journal}{Phys. Rev. D} \textbf{\bibinfo{volume}{80}},
  \bibinfo{pages}{124040} (\bibinfo{year}{2009}), \eprint{0912.1249}.

\bibitem[{\citenamefont{Harko}(2010)}]{Harko:2010zi}
\bibinfo{author}{\bibfnamefont{T.}~\bibnamefont{Harko}},
  \bibinfo{journal}{Phys. Rev. D} \textbf{\bibinfo{volume}{81}},
  \bibinfo{pages}{044021} (\bibinfo{year}{2010}), \eprint{1001.5349}.

\bibitem[{\citenamefont{Bertolami et~al.}(2010)\citenamefont{Bertolami, Frazao,
  and Paramos}}]{Bertolami:2010cw}
\bibinfo{author}{\bibfnamefont{O.}~\bibnamefont{Bertolami}},
  \bibinfo{author}{\bibfnamefont{P.}~\bibnamefont{Frazao}}, \bibnamefont{and}
  \bibinfo{author}{\bibfnamefont{J.}~\bibnamefont{Paramos}},
  \bibinfo{journal}{Phys. Rev.} \textbf{\bibinfo{volume}{D81}},
  \bibinfo{pages}{104046} (\bibinfo{year}{2010}), \eprint{1003.0850}.

\bibitem[{\citenamefont{Ribeiro and P{\'{a}}ramos}(2014)}]{Ribeiro2014}
\bibinfo{author}{\bibfnamefont{R.}~\bibnamefont{Ribeiro}} \bibnamefont{and}
  \bibinfo{author}{\bibfnamefont{J.}~\bibnamefont{P{\'{a}}ramos}},
  \bibinfo{journal}{Phys. Rev. D} \textbf{\bibinfo{volume}{90}},
  \bibinfo{pages}{124065} (\bibinfo{year}{2014}), \eprint{arXiv:1409.3046}.

\bibitem[{\citenamefont{Azizi and Yaraie}(2014)}]{Azizi:2014qsa}
\bibinfo{author}{\bibfnamefont{T.}~\bibnamefont{Azizi}} \bibnamefont{and}
  \bibinfo{author}{\bibfnamefont{E.}~\bibnamefont{Yaraie}},
  \bibinfo{journal}{Int. J. Mod. Phys. D} \textbf{\bibinfo{volume}{23}},
  \bibinfo{pages}{1450021} (\bibinfo{year}{2014}).

\bibitem[{\citenamefont{Bertolami and P{\'{a}}ramos}(2014)}]{Bertolami2014}
\bibinfo{author}{\bibfnamefont{O.}~\bibnamefont{Bertolami}} \bibnamefont{and}
  \bibinfo{author}{\bibfnamefont{J.}~\bibnamefont{P{\'{a}}ramos}},
  \bibinfo{journal}{Phys. Rev. D} \textbf{\bibinfo{volume}{89}},
  \bibinfo{pages}{044012} (\bibinfo{year}{2014}), \eprint{1311.5615}.

\bibitem[{\citenamefont{{Bekenstein}}(1982)}]{1982PhRvD..25.1527B}
\bibinfo{author}{\bibfnamefont{J.~D.} \bibnamefont{{Bekenstein}}},
  \bibinfo{journal}{\prd} \textbf{\bibinfo{volume}{25}}, \bibinfo{pages}{1527}
  (\bibinfo{year}{1982}).

\bibitem[{\citenamefont{Sandvik et~al.}(2002)\citenamefont{Sandvik, Barrow, and
  Magueijo}}]{Sandvik:2001rv}
\bibinfo{author}{\bibfnamefont{H.~B.} \bibnamefont{Sandvik}},
  \bibinfo{author}{\bibfnamefont{J.~D.} \bibnamefont{Barrow}},
  \bibnamefont{and} \bibinfo{author}{\bibfnamefont{J.}~\bibnamefont{Magueijo}},
  \bibinfo{journal}{Phys. Rev. Lett.} \textbf{\bibinfo{volume}{88}},
  \bibinfo{pages}{031302} (\bibinfo{year}{2002}), \eprint{astro-ph/0107512}.

\bibitem[{\citenamefont{Anchordoqui and Goldberg}(2003)}]{Anchordoqui:2003ij}
\bibinfo{author}{\bibfnamefont{L.}~\bibnamefont{Anchordoqui}} \bibnamefont{and}
  \bibinfo{author}{\bibfnamefont{H.}~\bibnamefont{Goldberg}},
  \bibinfo{journal}{Phys. Rev. D} \textbf{\bibinfo{volume}{68}},
  \bibinfo{pages}{083513} (\bibinfo{year}{2003}), \eprint{hep-ph/0306084}.

\bibitem[{\citenamefont{Copeland et~al.}(2004)\citenamefont{Copeland, Nunes,
  and Pospelov}}]{Copeland:2003cv}
\bibinfo{author}{\bibfnamefont{E.}~\bibnamefont{Copeland}},
  \bibinfo{author}{\bibfnamefont{N.}~\bibnamefont{Nunes}}, \bibnamefont{and}
  \bibinfo{author}{\bibfnamefont{M.}~\bibnamefont{Pospelov}},
  \bibinfo{journal}{Phys. Rev. D} \textbf{\bibinfo{volume}{69}},
  \bibinfo{pages}{023501} (\bibinfo{year}{2004}), \eprint{hep-ph/0307299}.

\bibitem[{\citenamefont{Lee et~al.}(2004)\citenamefont{Lee, Olive, and
  Pospelov}}]{Lee:2004vm}
\bibinfo{author}{\bibfnamefont{S.}~\bibnamefont{Lee}},
  \bibinfo{author}{\bibfnamefont{K.~A.} \bibnamefont{Olive}}, \bibnamefont{and}
  \bibinfo{author}{\bibfnamefont{M.}~\bibnamefont{Pospelov}},
  \bibinfo{journal}{Phys. Rev. D} \textbf{\bibinfo{volume}{70}},
  \bibinfo{pages}{083503} (\bibinfo{year}{2004}), \eprint{astro-ph/0406039}.

\bibitem[{\citenamefont{Koivisto}(2005)}]{Koivisto:2005nr}
\bibinfo{author}{\bibfnamefont{T.}~\bibnamefont{Koivisto}},
  \bibinfo{journal}{Phys. Rev. D} \textbf{\bibinfo{volume}{72}},
  \bibinfo{pages}{043516} (\bibinfo{year}{2005}), \eprint{astro-ph/0504571}.

\bibitem[{\citenamefont{Avelino}(2008)}]{Avelino:2008dc}
\bibinfo{author}{\bibfnamefont{P.~P.} \bibnamefont{Avelino}},
  \bibinfo{journal}{Phys. Rev. D} \textbf{\bibinfo{volume}{78}},
  \bibinfo{pages}{043516} (\bibinfo{year}{2008}), \eprint{0804.3394}.

\bibitem[{\citenamefont{Ayaita et~al.}(2012)\citenamefont{Ayaita, Weber, and
  Wetterich}}]{Ayata:2012}
\bibinfo{author}{\bibfnamefont{Y.}~\bibnamefont{Ayaita}},
  \bibinfo{author}{\bibfnamefont{M.}~\bibnamefont{Weber}}, \bibnamefont{and}
  \bibinfo{author}{\bibfnamefont{C.}~\bibnamefont{Wetterich}},
  \bibinfo{journal}{Phys. Rev. D} \textbf{\bibinfo{volume}{85}},
  \bibinfo{pages}{123010} (\bibinfo{year}{2012}).

\bibitem[{\citenamefont{Pourtsidou et~al.}(2013)\citenamefont{Pourtsidou,
  Skordis, and Copeland}}]{Pourtsidou:2013nha}
\bibinfo{author}{\bibfnamefont{A.}~\bibnamefont{Pourtsidou}},
  \bibinfo{author}{\bibfnamefont{C.}~\bibnamefont{Skordis}}, \bibnamefont{and}
  \bibinfo{author}{\bibfnamefont{E.}~\bibnamefont{Copeland}},
  \bibinfo{journal}{Phys. Rev. D} \textbf{\bibinfo{volume}{88}},
  \bibinfo{pages}{083505} (\bibinfo{year}{2013}), \eprint{1307.0458}.

\bibitem[{\citenamefont{Faraoni et~al.}(2014)\citenamefont{Faraoni, Dent, and
  Saridakis}}]{Faraoni:2014vra}
\bibinfo{author}{\bibfnamefont{V.}~\bibnamefont{Faraoni}},
  \bibinfo{author}{\bibfnamefont{J.~B.} \bibnamefont{Dent}}, \bibnamefont{and}
  \bibinfo{author}{\bibfnamefont{E.~N.} \bibnamefont{Saridakis}},
  \bibinfo{journal}{Phys. Rev. D} \textbf{\bibinfo{volume}{90}},
  \bibinfo{pages}{063510} (\bibinfo{year}{2014}), \eprint{1405.7288}.

\bibitem[{\citenamefont{Boehmer
  et~al.}(2015{\natexlab{a}})\citenamefont{Boehmer, Tamanini, and
  Wright}}]{Boehmer:2015kta}
\bibinfo{author}{\bibfnamefont{C.~G.} \bibnamefont{Boehmer}},
  \bibinfo{author}{\bibfnamefont{N.}~\bibnamefont{Tamanini}}, \bibnamefont{and}
  \bibinfo{author}{\bibfnamefont{M.}~\bibnamefont{Wright}},
  \bibinfo{journal}{Phys. Rev. D} \textbf{\bibinfo{volume}{91}},
  \bibinfo{pages}{123002} (\bibinfo{year}{2015}{\natexlab{a}}),
  \eprint{1501.06540}.

\bibitem[{\citenamefont{Boehmer
  et~al.}(2015{\natexlab{b}})\citenamefont{Boehmer, Tamanini, and
  Wright}}]{Boehmer:2015sha}
\bibinfo{author}{\bibfnamefont{C.~G.} \bibnamefont{Boehmer}},
  \bibinfo{author}{\bibfnamefont{N.}~\bibnamefont{Tamanini}}, \bibnamefont{and}
  \bibinfo{author}{\bibfnamefont{M.}~\bibnamefont{Wright}},
  \bibinfo{journal}{Phys. Rev. D} \textbf{\bibinfo{volume}{91}},
  \bibinfo{pages}{123003} (\bibinfo{year}{2015}{\natexlab{b}}),
  \eprint{1502.04030}.

\bibitem[{\citenamefont{Barros}(2019)}]{Barros:2019rdv}
\bibinfo{author}{\bibfnamefont{B.~J.} \bibnamefont{Barros}},
  \bibinfo{journal}{Phys. Rev. D} \textbf{\bibinfo{volume}{99}},
  \bibinfo{pages}{064051} (\bibinfo{year}{2019}), \eprint{1901.03972}.

\bibitem[{\citenamefont{Kase and Tsujikawa}(2020)}]{Kase:2019veo}
\bibinfo{author}{\bibfnamefont{R.}~\bibnamefont{Kase}} \bibnamefont{and}
  \bibinfo{author}{\bibfnamefont{S.}~\bibnamefont{Tsujikawa}},
  \bibinfo{journal}{Phys. Rev. D} \textbf{\bibinfo{volume}{101}},
  \bibinfo{pages}{063511} (\bibinfo{year}{2020}), \eprint{1910.02699}.

\bibitem[{\citenamefont{Azevedo and Avelino}(2018)}]{Azevedo2018a}
\bibinfo{author}{\bibfnamefont{R.~P.~L.} \bibnamefont{Azevedo}}
  \bibnamefont{and} \bibinfo{author}{\bibfnamefont{P.~P.}
  \bibnamefont{Avelino}}, \bibinfo{journal}{Phys. Rev. D}
  \textbf{\bibinfo{volume}{98}}, \bibinfo{pages}{064045}
  (\bibinfo{year}{2018}).

\bibitem[{\citenamefont{Azevedo and Avelino}(2019)}]{Azevedo2019a}
\bibinfo{author}{\bibfnamefont{R.~P.~L.} \bibnamefont{Azevedo}}
  \bibnamefont{and} \bibinfo{author}{\bibfnamefont{P.~P.}
  \bibnamefont{Avelino}}, \bibinfo{journal}{Phys. Rev. D}
  \textbf{\bibinfo{volume}{99}}, \bibinfo{pages}{064027}
  (\bibinfo{year}{2019}).

\bibitem[{\citenamefont{{Maartens}}(1996)}]{1996astro.ph..9119M}
\bibinfo{author}{\bibfnamefont{R.}~\bibnamefont{{Maartens}}},
  \bibinfo{journal}{arXiv e-prints} \bibinfo{eid}{astro-ph/9609119}
  (\bibinfo{year}{1996}), \eprint{astro-ph/9609119}.

\bibitem[{\citenamefont{Bettoni et~al.}(2011)\citenamefont{Bettoni, Liberati,
  and Sindoni}}]{Bettoni:2011fs}
\bibinfo{author}{\bibfnamefont{D.}~\bibnamefont{Bettoni}},
  \bibinfo{author}{\bibfnamefont{S.}~\bibnamefont{Liberati}}, \bibnamefont{and}
  \bibinfo{author}{\bibfnamefont{L.}~\bibnamefont{Sindoni}},
  \bibinfo{journal}{JCAP} \textbf{\bibinfo{volume}{11}}, \bibinfo{pages}{007}
  (\bibinfo{year}{2011}), \eprint{1108.1728}.

\bibitem[{\citenamefont{Bettoni and Liberati}(2015)}]{Bettoni:2015wla}
\bibinfo{author}{\bibfnamefont{D.}~\bibnamefont{Bettoni}} \bibnamefont{and}
  \bibinfo{author}{\bibfnamefont{S.}~\bibnamefont{Liberati}},
  \bibinfo{journal}{JCAP} \textbf{\bibinfo{volume}{08}}, \bibinfo{pages}{023}
  (\bibinfo{year}{2015}), \eprint{1502.06613}.

\bibitem[{\citenamefont{Dutta et~al.}(2017)\citenamefont{Dutta, Khyllep, and
  Tamanini}}]{PhysRevD.95.023515}
\bibinfo{author}{\bibfnamefont{J.}~\bibnamefont{Dutta}},
  \bibinfo{author}{\bibfnamefont{W.}~\bibnamefont{Khyllep}}, \bibnamefont{and}
  \bibinfo{author}{\bibfnamefont{N.}~\bibnamefont{Tamanini}},
  \bibinfo{journal}{Phys. Rev. D} \textbf{\bibinfo{volume}{95}},
  \bibinfo{pages}{023515} (\bibinfo{year}{2017}).

\bibitem[{\citenamefont{Koivisto et~al.}(2015)\citenamefont{Koivisto,
  Saridakis, and Tamanini}}]{Koivisto:2015qua}
\bibinfo{author}{\bibfnamefont{T.~S.} \bibnamefont{Koivisto}},
  \bibinfo{author}{\bibfnamefont{E.~N.} \bibnamefont{Saridakis}},
  \bibnamefont{and} \bibinfo{author}{\bibfnamefont{N.}~\bibnamefont{Tamanini}},
  \bibinfo{journal}{JCAP} \textbf{\bibinfo{volume}{09}}, \bibinfo{pages}{047}
  (\bibinfo{year}{2015}), \eprint{1505.07556}.

\bibitem[{\citenamefont{B\"ohmer et~al.}(2015)\citenamefont{B\"ohmer, Tamanini,
  and Wright}}]{PhysRevD.92.124067}
\bibinfo{author}{\bibfnamefont{C.~G.} \bibnamefont{B\"ohmer}},
  \bibinfo{author}{\bibfnamefont{N.}~\bibnamefont{Tamanini}}, \bibnamefont{and}
  \bibinfo{author}{\bibfnamefont{M.}~\bibnamefont{Wright}},
  \bibinfo{journal}{Phys. Rev. D} \textbf{\bibinfo{volume}{92}},
  \bibinfo{pages}{124067} (\bibinfo{year}{2015}).

\bibitem[{\citenamefont{Brax and Tamanini}(2016)}]{PhysRevD.93.103502}
\bibinfo{author}{\bibfnamefont{P.}~\bibnamefont{Brax}} \bibnamefont{and}
  \bibinfo{author}{\bibfnamefont{N.}~\bibnamefont{Tamanini}},
  \bibinfo{journal}{Phys. Rev. D} \textbf{\bibinfo{volume}{93}},
  \bibinfo{pages}{103502} (\bibinfo{year}{2016}).

\bibitem[{\citenamefont{Tamanini and Wright}(2016)}]{Tamanini:2016klr}
\bibinfo{author}{\bibfnamefont{N.}~\bibnamefont{Tamanini}} \bibnamefont{and}
  \bibinfo{author}{\bibfnamefont{M.}~\bibnamefont{Wright}},
  \bibinfo{journal}{JCAP} \textbf{\bibinfo{volume}{04}}, \bibinfo{pages}{032}
  (\bibinfo{year}{2016}), \eprint{1602.06903}.

\bibitem[{\citenamefont{Bailyn}(1980)}]{PhysRevD.22.267}
\bibinfo{author}{\bibfnamefont{M.}~\bibnamefont{Bailyn}},
  \bibinfo{journal}{Phys. Rev. D} \textbf{\bibinfo{volume}{22}},
  \bibinfo{pages}{267} (\bibinfo{year}{1980}).

\bibitem[{\citenamefont{Wongjun}(2017)}]{PhysRevD.96.023516}
\bibinfo{author}{\bibfnamefont{P.}~\bibnamefont{Wongjun}},
  \bibinfo{journal}{Phys. Rev. D} \textbf{\bibinfo{volume}{96}},
  \bibinfo{pages}{023516} (\bibinfo{year}{2017}).

\bibitem[{\citenamefont{{Diez-Tejedor}}(2013)}]{2013PhLB..727...27D}
\bibinfo{author}{\bibfnamefont{A.}~\bibnamefont{{Diez-Tejedor}}},
  \bibinfo{journal}{Physics Letters B} \textbf{\bibinfo{volume}{727}},
  \bibinfo{pages}{27} (\bibinfo{year}{2013}), \eprint{1309.4756}.

\bibitem[{\citenamefont{{Wetterich}}(1995)}]{1995A&A...301..321W}
\bibinfo{author}{\bibfnamefont{C.}~\bibnamefont{{Wetterich}}},
  \bibinfo{journal}{Astron. Astrophys.} \textbf{\bibinfo{volume}{301}},
  \bibinfo{pages}{321} (\bibinfo{year}{1995}), \eprint{hep-th/9408025}.

\bibitem[{\citenamefont{{Amendola}}(2000)}]{2000PhRvD..62d3511A}
\bibinfo{author}{\bibfnamefont{L.}~\bibnamefont{{Amendola}}},
  \bibinfo{journal}{\prd} \textbf{\bibinfo{volume}{62}}, \bibinfo{eid}{043511}
  (\bibinfo{year}{2000}), \eprint{astro-ph/9908023}.

\bibitem[{\citenamefont{{Zimdahl} et~al.}(2001)\citenamefont{{Zimdahl},
  {Pav{\'o}n}, and {Chimento}}}]{2001PhLB..521..133Z}
\bibinfo{author}{\bibfnamefont{W.}~\bibnamefont{{Zimdahl}}},
  \bibinfo{author}{\bibfnamefont{D.}~\bibnamefont{{Pav{\'o}n}}},
  \bibnamefont{and} \bibinfo{author}{\bibfnamefont{L.~P.}
  \bibnamefont{{Chimento}}}, \bibinfo{journal}{Physics Letters B}
  \textbf{\bibinfo{volume}{521}}, \bibinfo{pages}{133} (\bibinfo{year}{2001}),
  \eprint{astro-ph/0105479}.

\bibitem[{\citenamefont{{Farrar} and {Peebles}}(2004)}]{2004ApJ...604....1F}
\bibinfo{author}{\bibfnamefont{G.~R.} \bibnamefont{{Farrar}}} \bibnamefont{and}
  \bibinfo{author}{\bibfnamefont{P.~J.~E.} \bibnamefont{{Peebles}}},
  \bibinfo{journal}{\apj} \textbf{\bibinfo{volume}{604}}, \bibinfo{pages}{1}
  (\bibinfo{year}{2004}), \eprint{astro-ph/0307316}.

\bibitem[{\citenamefont{Brill and Wheeler}(1957)}]{RevModPhys.29.465}
\bibinfo{author}{\bibfnamefont{D.~R.} \bibnamefont{Brill}} \bibnamefont{and}
  \bibinfo{author}{\bibfnamefont{J.~A.} \bibnamefont{Wheeler}},
  \bibinfo{journal}{Rev. Mod. Phys.} \textbf{\bibinfo{volume}{29}},
  \bibinfo{pages}{465} (\bibinfo{year}{1957}).

\bibitem[{\citenamefont{Olive and Pospelov}(2002)}]{Olive:2001vz}
\bibinfo{author}{\bibfnamefont{K.~A.} \bibnamefont{Olive}} \bibnamefont{and}
  \bibinfo{author}{\bibfnamefont{M.}~\bibnamefont{Pospelov}},
  \bibinfo{journal}{Phys. Rev. D} \textbf{\bibinfo{volume}{65}},
  \bibinfo{pages}{085044} (\bibinfo{year}{2002}), \eprint{hep-ph/0110377}.

\bibitem[{\citenamefont{Avelino and Azevedo}(2020)}]{Avelino:2020fek}
\bibinfo{author}{\bibfnamefont{P.~P.} \bibnamefont{Avelino}} \bibnamefont{and}
  \bibinfo{author}{\bibfnamefont{R.~P.~L.} \bibnamefont{Azevedo}}
  (\bibinfo{year}{2020}), \eprint{2003.10154}.

\bibitem[{\citenamefont{Avelino
  et~al.}(2008{\natexlab{a}})\citenamefont{Avelino, Beca, and
  Martins}}]{Avelino:2007tu}
\bibinfo{author}{\bibfnamefont{P.~P.} \bibnamefont{Avelino}},
  \bibinfo{author}{\bibfnamefont{L.~M.~G.} \bibnamefont{Beca}},
  \bibnamefont{and} \bibinfo{author}{\bibfnamefont{C.~J. A.~P.}
  \bibnamefont{Martins}}, \bibinfo{journal}{Phys. Rev. D}
  \textbf{\bibinfo{volume}{77}}, \bibinfo{pages}{063515}
  (\bibinfo{year}{2008}{\natexlab{a}}), \eprint{0711.4288}.

\bibitem[{\citenamefont{Avelino et~al.}(2003)\citenamefont{Avelino, Beca,
  de~Carvalho, and Martins}}]{Avelino:2003cf}
\bibinfo{author}{\bibfnamefont{P.~P.} \bibnamefont{Avelino}},
  \bibinfo{author}{\bibfnamefont{L.~M.~G.} \bibnamefont{Beca}},
  \bibinfo{author}{\bibfnamefont{J.~P.~M.} \bibnamefont{de~Carvalho}},
  \bibnamefont{and} \bibinfo{author}{\bibfnamefont{C.~J. A.~P.}
  \bibnamefont{Martins}}, \bibinfo{journal}{JCAP}
  \textbf{\bibinfo{volume}{09}}, \bibinfo{pages}{002} (\bibinfo{year}{2003}),
  \eprint{astro-ph/0307427}.

\bibitem[{\citenamefont{{Mukhanov}}(2005)}]{Mukhanov:2005sc}
\bibinfo{author}{\bibfnamefont{V.}~\bibnamefont{{Mukhanov}}},
  \emph{\bibinfo{title}{{Physical Foundations of Cosmology}}}
  (\bibinfo{publisher}{Cambridge University Press}, \bibinfo{address}{Oxford},
  \bibinfo{year}{2005}).

\bibitem[{\citenamefont{Avelino
  et~al.}(2008{\natexlab{b}})\citenamefont{Avelino, Beca, and
  Martins}}]{Avelino:2008cu}
\bibinfo{author}{\bibfnamefont{P.~P.} \bibnamefont{Avelino}},
  \bibinfo{author}{\bibfnamefont{L.~M.~G.} \bibnamefont{Beca}},
  \bibnamefont{and} \bibinfo{author}{\bibfnamefont{C.~J. A.~P.}
  \bibnamefont{Martins}}, \bibinfo{journal}{Phys. Rev. D}
  \textbf{\bibinfo{volume}{77}}, \bibinfo{pages}{101302}
  (\bibinfo{year}{2008}{\natexlab{b}}), \eprint{0802.0174}.

\bibitem[{\citenamefont{{Chimento}}(2004)}]{2004PhRvD..69l3517C}
\bibinfo{author}{\bibfnamefont{L.~P.} \bibnamefont{{Chimento}}},
  \bibinfo{journal}{Physical Review D} \textbf{\bibinfo{volume}{69}},
  \bibinfo{eid}{123517} (\bibinfo{year}{2004}), \eprint{astro-ph/0311613}.

\bibitem[{\citenamefont{{Garriga} and {Mukhanov}}(1999)}]{1999PhLB..458..219G}
\bibinfo{author}{\bibfnamefont{J.}~\bibnamefont{{Garriga}}} \bibnamefont{and}
  \bibinfo{author}{\bibfnamefont{V.~F.} \bibnamefont{{Mukhanov}}},
  \bibinfo{journal}{Physics Letters B} \textbf{\bibinfo{volume}{458}},
  \bibinfo{pages}{219} (\bibinfo{year}{1999}), \eprint{hep-th/9904176}.

\bibitem[{\citenamefont{Bartolo et~al.}(2004)\citenamefont{Bartolo, Corasaniti,
  Liddle, and Malquarti}}]{Bartolo:2003ad}
\bibinfo{author}{\bibfnamefont{N.}~\bibnamefont{Bartolo}},
  \bibinfo{author}{\bibfnamefont{P.~S.} \bibnamefont{Corasaniti}},
  \bibinfo{author}{\bibfnamefont{A.~R.} \bibnamefont{Liddle}},
  \bibnamefont{and}
  \bibinfo{author}{\bibfnamefont{M.}~\bibnamefont{Malquarti}},
  \bibinfo{journal}{Phys. Rev. D} \textbf{\bibinfo{volume}{70}},
  \bibinfo{pages}{043532} (\bibinfo{year}{2004}), \eprint{astro-ph/0311503}.

\bibitem[{\citenamefont{Bamba et~al.}(2012{\natexlab{b}})\citenamefont{Bamba,
  Matsumoto, and Nojiri}}]{Bamba:2011ih}
\bibinfo{author}{\bibfnamefont{K.}~\bibnamefont{Bamba}},
  \bibinfo{author}{\bibfnamefont{J.}~\bibnamefont{Matsumoto}},
  \bibnamefont{and} \bibinfo{author}{\bibfnamefont{S.}~\bibnamefont{Nojiri}},
  \bibinfo{journal}{Phys. Rev. D} \textbf{\bibinfo{volume}{85}},
  \bibinfo{pages}{084026} (\bibinfo{year}{2012}{\natexlab{b}}),
  \eprint{1109.1308}.

\bibitem[{\citenamefont{{Linder}}(2003)}]{2003PhRvL..90i1301L}
\bibinfo{author}{\bibfnamefont{E.~V.} \bibnamefont{{Linder}}},
  \bibinfo{journal}{\prl} \textbf{\bibinfo{volume}{90}}, \bibinfo{eid}{091301}
  (\bibinfo{year}{2003}), \eprint{astro-ph/0208512}.

\bibitem[{\citenamefont{{de Putter} and {Linder}}(2007)}]{2007APh....28..263D}
\bibinfo{author}{\bibfnamefont{R.}~\bibnamefont{{de Putter}}} \bibnamefont{and}
  \bibinfo{author}{\bibfnamefont{E.~V.} \bibnamefont{{Linder}}},
  \bibinfo{journal}{Astroparticle Physics} \textbf{\bibinfo{volume}{28}},
  \bibinfo{pages}{263} (\bibinfo{year}{2007}), \eprint{0705.0400}.

\bibitem[{\citenamefont{{Ferreira} and {Avelino}}(2017)}]{2017PhLB..770..213F}
\bibinfo{author}{\bibfnamefont{V.~M.~C.} \bibnamefont{{Ferreira}}}
  \bibnamefont{and} \bibinfo{author}{\bibfnamefont{P.~P.}
  \bibnamefont{{Avelino}}}, \bibinfo{journal}{Physics Letters B}
  \textbf{\bibinfo{volume}{770}}, \bibinfo{pages}{213} (\bibinfo{year}{2017}),
  \eprint{1611.08403}.

\bibitem[{\citenamefont{{Sandvik} et~al.}(2004)\citenamefont{{Sandvik},
  {Tegmark}, {Zaldarriaga}, and {Waga}}}]{2004PhRvD..69l3524S}
\bibinfo{author}{\bibfnamefont{H.~B.} \bibnamefont{{Sandvik}}},
  \bibinfo{author}{\bibfnamefont{M.}~\bibnamefont{{Tegmark}}},
  \bibinfo{author}{\bibfnamefont{M.}~\bibnamefont{{Zaldarriaga}}},
  \bibnamefont{and} \bibinfo{author}{\bibfnamefont{I.}~\bibnamefont{{Waga}}},
  \bibinfo{journal}{\prd} \textbf{\bibinfo{volume}{69}}, \bibinfo{eid}{123524}
  (\bibinfo{year}{2004}), \eprint{astro-ph/0212114}.

\bibitem[{\citenamefont{Avelino and Ferreira}(2015)}]{Avelino:2015dwa}
\bibinfo{author}{\bibfnamefont{P.~P.} \bibnamefont{Avelino}} \bibnamefont{and}
  \bibinfo{author}{\bibfnamefont{V.~M.~C.} \bibnamefont{Ferreira}},
  \bibinfo{journal}{Phys. Rev. D} \textbf{\bibinfo{volume}{91}},
  \bibinfo{pages}{083508} (\bibinfo{year}{2015}), \eprint{1502.07583}.

\bibitem[{\citenamefont{{Avelino} et~al.}(2004)\citenamefont{{Avelino},
  {Be{\c{c}}a}, {de Carvalho}, {Martins}, and
  {Copeland}}}]{2004PhRvD..69d1301A}
\bibinfo{author}{\bibfnamefont{P.~P.} \bibnamefont{{Avelino}}},
  \bibinfo{author}{\bibfnamefont{L.~M.} \bibnamefont{{Be{\c{c}}a}}},
  \bibinfo{author}{\bibfnamefont{J.~P.} \bibnamefont{{de Carvalho}}},
  \bibinfo{author}{\bibfnamefont{C.~J.} \bibnamefont{{Martins}}},
  \bibnamefont{and} \bibinfo{author}{\bibfnamefont{E.~J.}
  \bibnamefont{{Copeland}}}, \bibinfo{journal}{\prd}
  \textbf{\bibinfo{volume}{69}}, \bibinfo{eid}{041301} (\bibinfo{year}{2004}),
  \eprint{astro-ph/0306493}.

\bibitem[{\citenamefont{{Bilic} et~al.}(2004)\citenamefont{{Bilic},
  {Lindebaum}, {Tupper}, and {Viollier}}}]{2004JCAP...11..008B}
\bibinfo{author}{\bibfnamefont{N.}~\bibnamefont{{Bilic}}},
  \bibinfo{author}{\bibfnamefont{R.~J.} \bibnamefont{{Lindebaum}}},
  \bibinfo{author}{\bibfnamefont{G.~B.} \bibnamefont{{Tupper}}},
  \bibnamefont{and} \bibinfo{author}{\bibfnamefont{R.~D.}
  \bibnamefont{{Viollier}}}, \bibinfo{journal}{\jcap}
  \textbf{\bibinfo{volume}{2004}}, \bibinfo{eid}{008} (\bibinfo{year}{2004}),
  \eprint{astro-ph/0307214}.

\bibitem[{\citenamefont{{Roy} and {Buchert}}(2010)}]{2010CQGra..27q5013R}
\bibinfo{author}{\bibfnamefont{X.}~\bibnamefont{{Roy}}} \bibnamefont{and}
  \bibinfo{author}{\bibfnamefont{T.}~\bibnamefont{{Buchert}}},
  \bibinfo{journal}{Classical and Quantum Gravity}
  \textbf{\bibinfo{volume}{27}}, \bibinfo{eid}{175013} (\bibinfo{year}{2010}),
  \eprint{0909.4155}.

\bibitem[{\citenamefont{Beca and Avelino}(2007)}]{Beca:2005gc}
\bibinfo{author}{\bibfnamefont{L.~M.~G.} \bibnamefont{Beca}} \bibnamefont{and}
  \bibinfo{author}{\bibfnamefont{P.~P.} \bibnamefont{Avelino}},
  \bibinfo{journal}{Mon. Not. Roy. Astron. Soc.}
  \textbf{\bibinfo{volume}{376}}, \bibinfo{pages}{1169} (\bibinfo{year}{2007}),
  \eprint{astro-ph/0507075}.

\bibitem[{\citenamefont{{Del Popolo} et~al.}(2013)\citenamefont{{Del Popolo},
  {Pace}, {Maydanyuk}, {Lima}, and {Jesus}}}]{2013PhRvD..87d3527D}
\bibinfo{author}{\bibfnamefont{A.}~\bibnamefont{{Del Popolo}}},
  \bibinfo{author}{\bibfnamefont{F.}~\bibnamefont{{Pace}}},
  \bibinfo{author}{\bibfnamefont{S.~P.} \bibnamefont{{Maydanyuk}}},
  \bibinfo{author}{\bibfnamefont{J.~A.~S.} \bibnamefont{{Lima}}},
  \bibnamefont{and} \bibinfo{author}{\bibfnamefont{J.~F.}
  \bibnamefont{{Jesus}}}, \bibinfo{journal}{\prd}
  \textbf{\bibinfo{volume}{87}}, \bibinfo{eid}{043527} (\bibinfo{year}{2013}),
  \eprint{1303.3628}.

\bibitem[{\citenamefont{{Avelino} et~al.}(2014)\citenamefont{{Avelino},
  {Bolejko}, and {Lewis}}}]{2014PhRvD..89j3004A}
\bibinfo{author}{\bibfnamefont{P.~P.} \bibnamefont{{Avelino}}},
  \bibinfo{author}{\bibfnamefont{K.}~\bibnamefont{{Bolejko}}},
  \bibnamefont{and} \bibinfo{author}{\bibfnamefont{G.~F.}
  \bibnamefont{{Lewis}}}, \bibinfo{journal}{\prd}
  \textbf{\bibinfo{volume}{89}}, \bibinfo{eid}{103004} (\bibinfo{year}{2014}),
  \eprint{1403.1718}.

\end{thebibliography}

\end{document}